\documentclass[8.5pt,twoside,twocolumn]{article}
\usepackage[super,sort&compress,comma]{natbib} 
\usepackage{natmove}
\usepackage{times,mathptmx}
\usepackage{graphicx} 
\usepackage{lastpage}
\usepackage[format=plain,justification=raggedright,singlelinecheck=false,font=small,labelfont=bf,labelsep=space]{caption} 

\usepackage{secsty} 
\usepackage{balance} 
\usepackage{fancyhdr}

\newlength{\narrowf}
\newlength{\widef}
\newlength{\littlepic}
\begin{document}

\thispagestyle{plain}
\fancypagestyle{plain}{
\renewcommand{\headrulewidth}{1pt}}
\renewcommand{\thefootnote}{\fnsymbol{footnote}}
\renewcommand\footnoterule{\vspace*{1pt}%
\hrule width 3.4in height 0.4pt \vspace*{5pt}} 
\setcounter{secnumdepth}{5}

\makeatletter 
\def\subsubsection{\@startsection{subsubsection}{3}{10pt}{-1.25ex plus -1ex minus -.1ex}{0ex plus 0ex}{\normalsize\bf}} 
\def\paragraph{\@startsection{paragraph}{4}{10pt}{-1.25ex plus -1ex minus -.1ex}{0ex plus 0ex}{\normalsize\textit}} 
\renewcommand\@biblabel[1]{#1}            
\renewcommand\@makefntext[1]%
{\noindent\makebox[0pt][r]{\@thefnmark\,}#1}
\makeatother 
\renewcommand{\figurename}{\small{Fig.}~}
\sectionfont{\large}
\subsectionfont{\normalsize} 

\fancyfoot{}
\fancyfoot[RO]{\footnotesize{\sffamily{\thepage}}}
\fancyfoot[LE]{\footnotesize{\sffamily{\thepage}}}
\fancyhead{}
\renewcommand{\headrulewidth}{1pt} 
\renewcommand{\footrulewidth}{1pt}
\setlength{\arrayrulewidth}{1pt}
\setlength{\columnsep}{6.5mm}
\setlength\bibsep{1pt}



\twocolumn[
  \begin{@twocolumnfalse}
\noindent\LARGE{\textbf{Particle Shape Effects on the Stress Response of Granular Packings}}
\vspace{0.6cm}

\noindent\large{\textbf{Athanasios G. Athanassiadis,\textit{$^{a}$} Marc Z. Miskin,\textit{$^{a}$} Paul Kaplan,\textit{$^{a}$} Nicholas Rodenberg,\textit{$^{a}$} Seung Hwan Lee,\textit{$^{a}$} Jason Merritt,\textit{$^{a}$} Eric Brown,\textit{$^{a}$} John Amend,\textit{$^{b}$} Hod Lipson,\textit{$^{b}$} and
Heinrich M. Jaeger$^{\ast}$\textit{$^{a}$}}}\vspace{0.5cm}

\vspace{1.1cm}

\noindent\normalsize{ We present measurements of the stress response of packings formed from a wide range of particle shapes. Besides spheres these include convex shapes such as the Platonic solids, truncated tetrahedra, and triangular bipyramids, as well as more complex, non-convex geometries such as hexapods with various arm lengths, dolos, and tetrahedral frames. All particles were 3D-printed in hard resin. Well-defined initial packing states were established through preconditioning by cyclic loading under given confinement pressure. Starting from such initial states, stress-strain relationships for axial compression were obtained at four different confining pressures for each particle type.  While confining pressure has the largest overall effect on the mechanical response, we find that particle shape controls the details of the stress-strain curves and can be used to tune  packing  stiffness and yielding. By correlating the experimentally measured values for the effective Young's modulus under compression, yield stress  and  energy loss during cyclic loading, we identify trends among the various shapes that allow for designing a packing's aggregate behavior.}
\vspace{1.5cm}
 \end{@twocolumnfalse}
  ]



    \footnotetext{\textit{$^{a}$~James Franck Institute \& Department of Physics, The University of Chicago, 929 East 57th Street, Chicago, IL 60637, USA; E-mail: h-jaeger@uchicago.edu}}
    \footnotetext{\textit{$^{b}$~Sibley School of Mechanical \& Aerospace Engineering, Cornell University, 105 Upson Hall, Ithaca, NY 14853, USA.}}

	\section*{Introduction}
    
    One of the fundamental challenges for granular physics is to identify links between properties of  individual particles and the resulting overall behavior observed when these particles are randomly packed into large aggregates. While it has long been recognized that particle shape plays a significant role in controlling a granular material's microstructure,\cite{CookeSegregation, WoodSoilMech, JaegerGranSLG, DuranGran, PoschelCompGran} most work to date using three-dimensional particles has focused on spheres and a small set of anisotropic shapes, such as ellipsoids,\cite{BuchalterEllipses, DonevUnderconstr, DonevDensEllipsoids, Schaller1} and rods.\cite{WouterseContactN, DesmondProlate} Recently, progress has been made by systematically investigating the microstructural configurations of more complex shapes including faceted polyhedra,\cite{WouterseShape, HajiAkbariTets, TorqPlatArch, BakerPlat, DelaneySuperellipsoids, JaoshTets, NeudeckerTets, JiaGeometry, TorqHardPart, KyrylyukShape, ZhaoShape, ShepCubes, TorqNonsph, FraigeCubes, MalinSpiky, SalotDEMTriax} often with the particular goal of finding the highest achievable packing fraction.\cite{HajiAkbariTets, TorqPlatArch, BakerPlat, JaoshTets, TorqHardPart, TorqNonsph} By contrast, the response of aggregates of non-spherical particles to applied mechanical loads has been explored much less.\cite{DesmondProlate, SalotDEMTriax, MatuttisDEM, SchreckDimer, HartlShape, Azema2013, Azema2013b} Furthermore, the vast majority of work so far has concentrated on convex particle shapes. Non-convex shapes can support types of contacts that make it possible for neighboring particles to interact in completely different ways such as by interlocking or entanglement.\cite{MalinSpiky, SchreckDimer, GravishEntangle, BrownChains, LudewigInterlocking, MengSpheroCyl}
    
	Shape-mediated particle interactions lead to opportunities to generate granular materials with special properties. Generally, as more complex geometries are explored, the packing's behavior is dependent not only upon the \textit{number} of local contacts, but also upon the geometrically-determined \textit{types} of contacts. For example, certain faceted polyhedra pack into particularly dense aggregates,\cite{HajiAkbariTets, TorqPlatArch, BakerPlat, JaoshTets, ShepCubes, TorqNonsph} while random packings of non-convex particles generically exhibit a much higher porosity.\cite{MalinSpiky, LudewigInterlocking, MengSpheroCyl, RemondNonConvex, ZouChains} With spheres the average number of local contacts controls the mechanical response, and we can expect that denser packings of the same spheres will be stiffer.\cite{TorqHardPart, SchreckDimer, LiuJamming, SomfaiCritJam, vanHeckeJam} With non-convex shapes this no longer has to be the case, making it possible to envision highly porous packings that nevertheless excel in stiffness. Therefore, for particles that are able to interlock or entangle, low packing fraction does not have to be incongruous with a high degree of mechanical stability. 
	
	This opens up a vast new portion of response space controlled by particle shape. If understood properly,  shape can be employed to design unique granular behaviors in novel applications that require carefully tuned or optimized aggregate properties. Among the newest of these are applications of granular materials in the fabrication of shapeable molds (`vacuumatics'),\cite{HuijbenVac2, SchmidtVac, HuijbenVac1} in `aggregate architecture,'\cite{DierichsMatComp} and in jamming-based soft robotics.\cite{BrownGripper, SteltzSoftRob, AmendGripper}
    
    However, several difficulties arise when dealing with non-spherical shapes. To begin with, contacts no longer are all of the same type. For example, faceted polyhedra produce different local interactions depending on whether one is dealing with face-face, face-edge or edge-edge contacts. In simulations of the stress response, this brings up questions regarding the proper contact force law for each of these cases. One way around this issue has been to model complex shapes as particles composed of rigidly connected, overlapping spheres or ellipsoids,\cite{PoschelCompGran, MalinSpiky, SalotDEMTriax, SchreckDimer, HartlShape, KodamDEM, MiskinArtEv} but we can expect that in many circumstances faceted particles will behave differently.  On the experimental side, one general limitation has been that the set of three-dimensional particle shapes available for testing was confined to either naturally occurring sands or soils, commercially available particle types,\cite{BakerPlat, JaoshTets} or particles made with special molds.\cite{NeudeckerTets} Advances such as three-dimensional rapid prototyping (3D-printing) have only become sufficiently accessible in the last few years to allow for the fabrication of arbitrarily shaped particles in sizes and surface finish suitable for granular materials testing. As a result, there have been no systematic investigations of how the mechanical response of granular packings changes when particle shape is varied across a wide range of convex and non-convex geometries. 
	
	The purpose of this paper is to fill this gap and provide base-line data. Using high-resolution 3D-printing, we fabricated sets of 14 different particle shapes. Eight of these shapes were convex, including the sphere, all the Platonic solids (tetrahedron, octahedron, cube, icosahedron, and dodecahedron), the truncated tetrahedron (an Archimedean solid), and the triangular bipyramid (a Johnson solid). The remaining 6 shapes were non-convex: tetrahedral frames, which can interpenetrate because of their open interior, hexapods (`jacks' consisting of a central sphere with six radial arms shaped as truncated cones) and dolos (H-shaped particles with one of the vertical arms rotated 90 degrees out of the plane; cast several meters tall from concrete and weighing in excess of 20 tons each, hexapods and dolos are typical particle shapes used for the outer layer of breakwaters, where interlocking helps to reduce particle displacement due to wave action\cite{BurcharthDolosse, GurerTetrapod}).  
    
    \begin{figure}[!t]
        \centering
		\includegraphics[width=.95\narrowf]{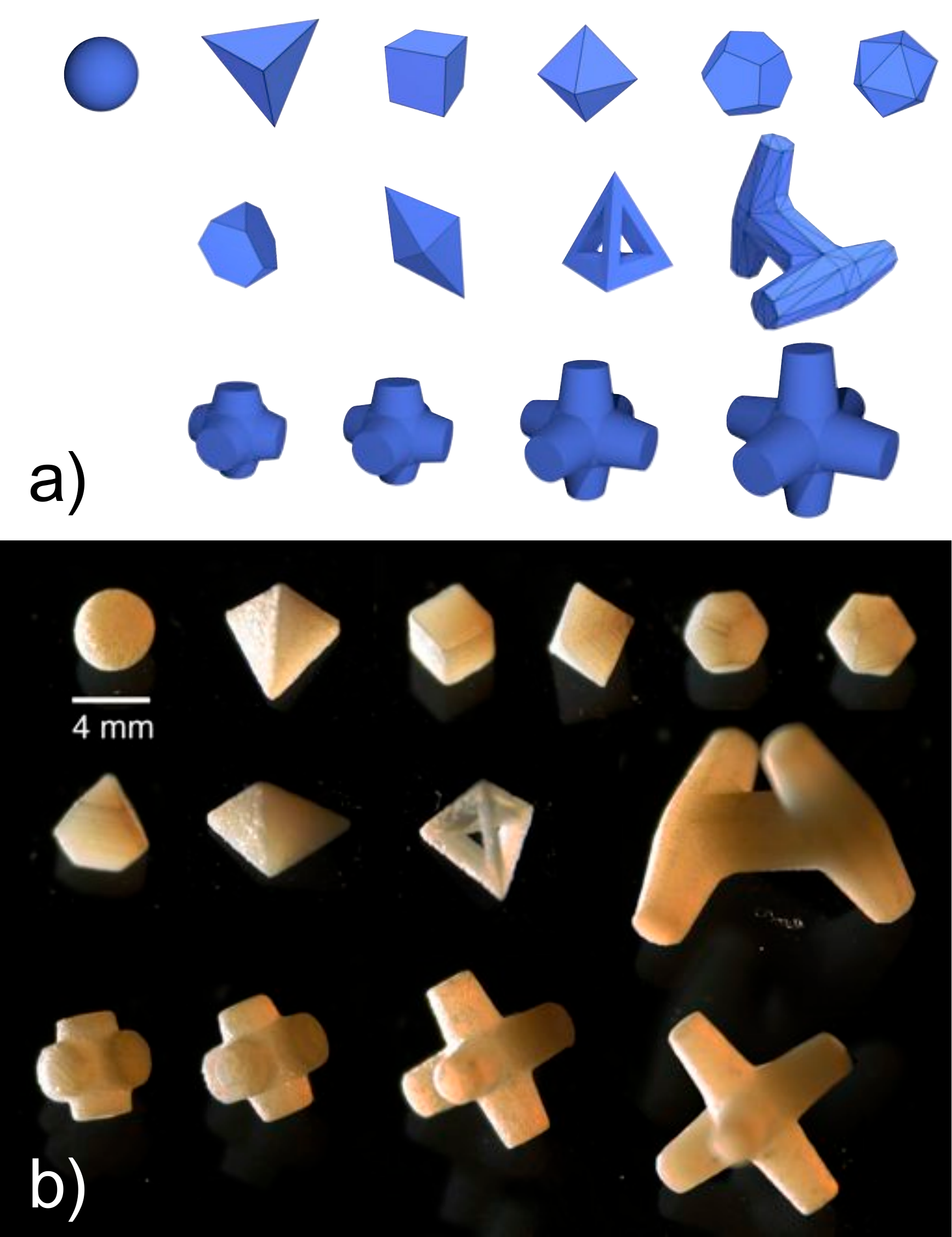}\\
		\caption{Particle Geometries. \textbf{(a)}~Computer renderings (not to scale) and \textbf{(b)}~photo of 3D-printed particles, after some use in the experiments. Top row (left to right): sphere, tetrahedron, cube, octahedron, dodecahedron, icosahedron. Middle row (left to right): truncated tetrahedron, triangular bipyramid, tetrahedral frame, dolo. Bottom row: jacks with arm length increasing to the right.}
        \label{fig:Grains}
	\end{figure}
	
	By measuring the stress-strain relationship of the particle aggregate under quasistatic compression, we focus here  on the overall, macroscopic response. For each particle type, we performed three to five triaxial tests at four different confining pressures for a total of 190 independent experiments. From our data we extract parameters characterizing the aggregate performance for each shape, such as an effective Young's modulus, a yield stress, and the amount of energy loss during cyclic compression. While the data from our experiments cannot reveal specific microstructural (re-)configurations during loading, it does provide an extensive overview of trends that emerge when shape is varied. Further, we are able to identify  correlations among the aggregate parameters and how they vary with geometric characteristics of the particle shape. These experimental data provide both a benchmark for comparison with simulations and a reference guide for picking appropriate shapes for applications.

	\section*{Materials and Experimental Procedure}

	Fig.~\ref{fig:Grains}a shows renderings of the 14 geometric models used to 3D-print the particles. We designed all eight convex shapes to have equal volume $V=$22.5~mm${}^3$, corresponding to a side length of 2.8~mm for the cubes. The tetrahedral frames have the same outer dimensions as the solid tetrahedron, with beams along the edges thick enough to withstand the stresses in the experiments without breaking (1.4~mm). The four jacks are formed by a central sphere with six truncated, conical arms at right angles; the only parameter we varied was the arm length (0.92~mm, 1.3~mm, 2.6~mm, 3.6~mm). The dolos (twisted `H') were printed at a larger size {(10.7~mm arm length)}. Additional geometry information is available in the Appendix. We have also made the models freely available for viewing and download.\cite{JaegerGrainSite}

	\begin{figure}
        \centering
		\includegraphics[width=\narrowf]{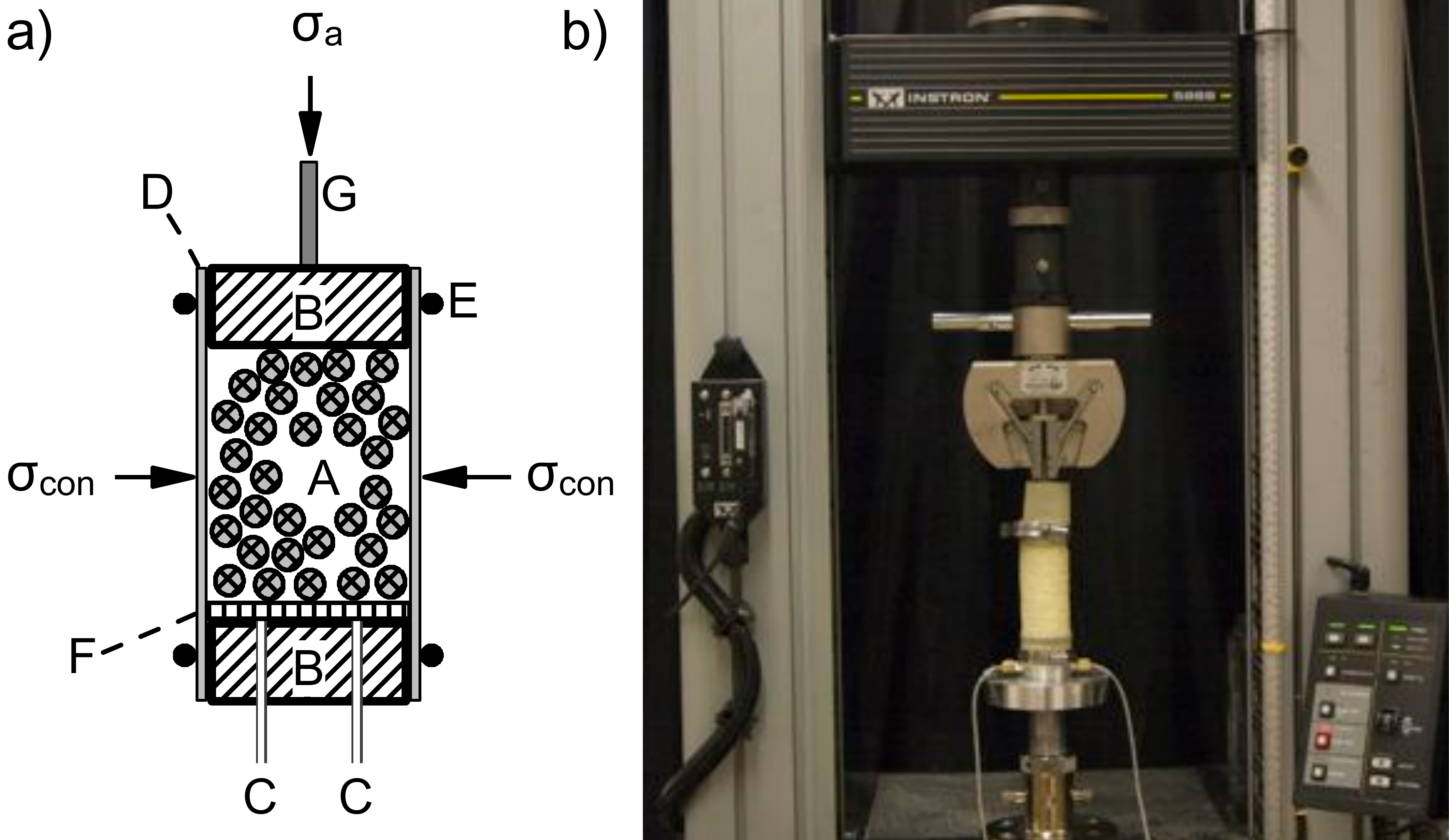}
		\caption{Experimental set-up. \textbf{(a)}~ Schematic of the triaxial test used to measure the mechanical response, with radial confining pressure $\sigma_{con}$ and axial pressure $\sigma_a = q + \sigma_{con}$, where $q$ is the applied deviatoric stress. A~-~granular aggregate; B~-~aluminum end caps; C~-~lines to vacuum pump and pressure gauge; D~-~thin latex membrane; E~-~rubber o-rings; F~-~porous disk; G~-~loading piston. \textbf{(b)}~Image of setup, with the jaws of the Instron materials tester connected to the loading piston (G). The granular packing (A) inside the semi-translucent membrane (D) appears white.}
        \label{fig:Setup}
	\end{figure}

	We printed the particles in sets of ${\sim5500}$ on an Objet Connex 350 3D-printer, using $50\mu$m print resolution and a UV-cured resin (``Vero~White~Plus'', Objet Geometries Inc.). To characterize the resin material itself, we compressed individual cubes to determine a compressive modulus $E_{mat}=1260\pm120$~MPa and measured an angle of maximum stability $\theta$=26$\pm$3$^\circ$ for spheres (see Table~\ref{tab:phi} for other shapes). During the printing process, the particles were embedded in a waxy support material that needed to be cleaned off thoroughly before assembling the packings. We cleaned the particles by crumbling off large chunks of support material by hand, and then placing the particles in 10\%~(by volume) NaOH solution for 1.5~hr while agitating with a magnetic stirrer. Afterward, residual NaOH and support material were removed with a high pressure water jet and the particles dried in air. Fig.~\ref{fig:Grains}b shows the printed particles after cleaning and some use in experiments.
    
	\begin{figure}
        \centering
		\includegraphics[width=\narrowf]{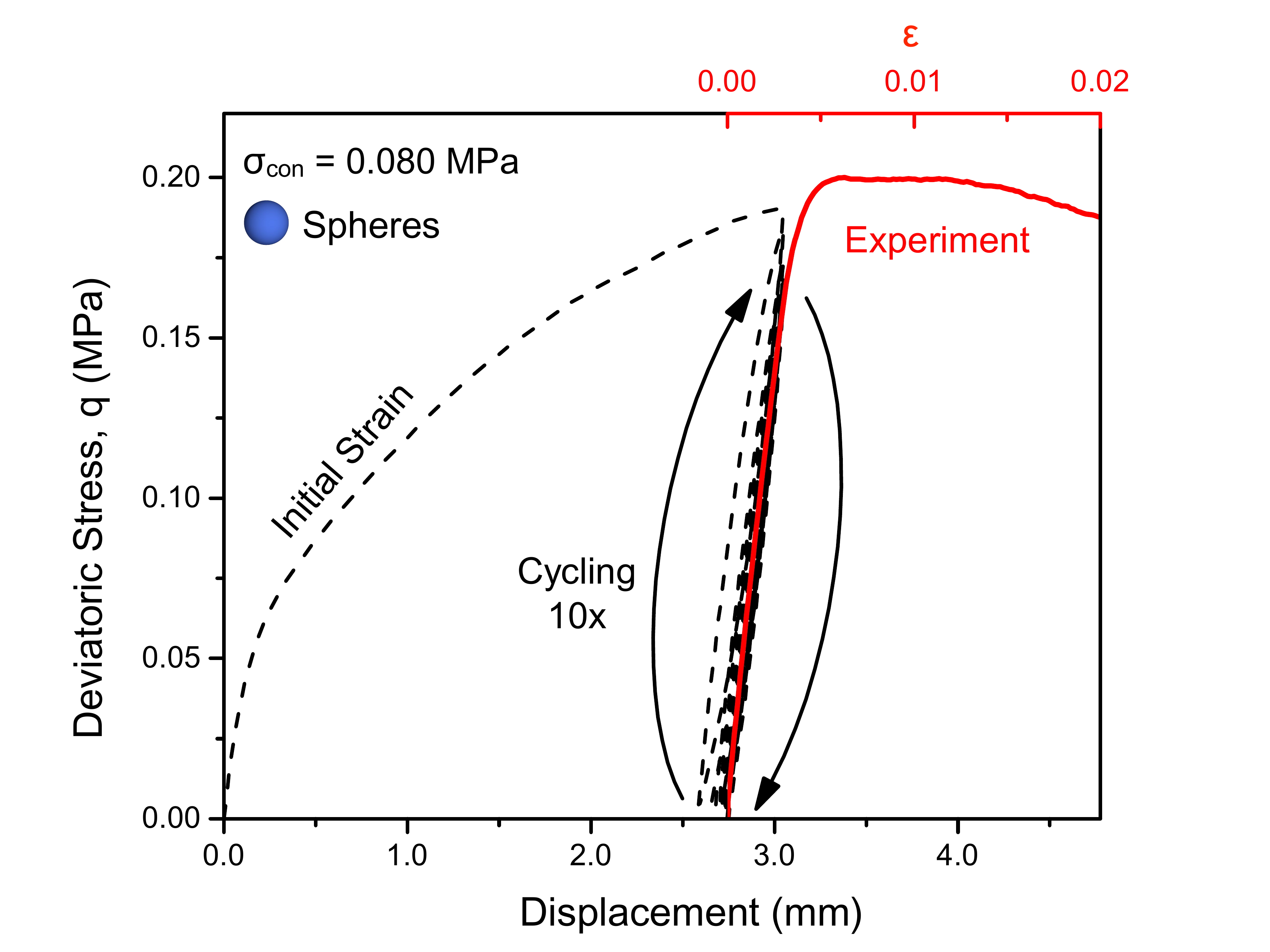}
		\caption{Initial stress-strain curve and conditioning by cyclic loading. This stress-strain curve for spheres at $\sigma_{con}$ = 0.080~MPa shows the initial compression and the 10 conditioning cycles (black dotted line), as well as the final compression run past failure (red solid line). The initial compression starts from the  isotropic stress state and axially strains the packing up to 3\% (3mm displacement). After 10 unloading/reloading cycles up to the same maximum displacement, we define the state with deviatoric stress $q$ = 0 as the conditioned reference state with $\varepsilon$=0 (red scale above figure).}
    \label{fig:CycleEx}
	\end{figure}
	
	For the mechanical tests, we measured the stress response of the granular packings in a high-precision triaxial test. We prepared random packings by slowly pouring particles through a funnel into a cylindrical latex membrane (Durham Geo-Enterprises, 0.30~mm thickness) of inner diameter $d$=50.8~mm, filling it to a height $h$=102~mm (Fig.~\ref{fig:Setup}a). In this configuration, each sample contained {5000-5500} particles and measured 15-20 particles across the membrane diameter (except dolos and large jacks, which had closer to 10 across). We maintained this $d$:$h$ = 1:2 initial sample aspect ratio for all triaxial tests, as is standard in soil mechanics.\cite{BishopTriax} Once the sample was loaded, we capped the open top of the membrane with an aluminum disc and rigidly connected the sample to the testing apparatus, an Instron 5869 materials tester (Fig.~\ref{fig:Setup}b). The bottom end cap of the packing was covered by a porous sintered disc and was connected to a vacuum pump. This pump allowed us to apply confining pressures between 0.001~MPa and 0.080~MPa to the packing during the triaxial test.
	
	To guarantee reproducibility in triaxial compression tests with frictional particles, care must be taken ensure a uniform, isotropic confining stress prior to any additional axial compression. We achieved this with the following protocol. We first applied the desired radial confining stress $\sigma_{con}$ to the sample using vacuum, while the top cap was rigidly held in place by the material tester. Monitoring the axial stress $\sigma_{a}$ with the Instron's load cell, we then lowered the top cap until it matched the radial confining stress.
	
    This stress-balanced state defined our initial state with deviatoric stress $q = \sigma_{a} - \sigma_{con}=0$. Next, to reduce effects from run-to-run variations associated with sample preparation, we axially compressed each sample by ~3\% (vertical displacement of ~3mm for sample height 102~mm) and then returned to $q$=0. The area enclosed by this loading/unloading curve provides a direct measure of the energy lost to friction and local rearrangements when the initial packing, after having been poured and confined, is consolidated for the first time. As seen in Fig.~\ref{fig:CycleEx}, repeated cycling up to the same maximum vertical displacement produces a set of hysteresis loops.

	In our experiments, we found that conditioning the packing with $N$ = 10 cycles resulted in a state that was largely independent of the initial pouring and yielded stress-strain curves that were highly reproducible from run to run (we discuss this in more detail below, see Fig. 7). We therefore used the state with $q$ = 0 after $N$ = 10 cycles as our reference state, resetting $\varepsilon$ = 0 (red scale in Fig.~\ref{fig:CycleEx}). To map out the mechanical response, packings were further compressed up to $\varepsilon=0.02$, which includes the regime beyond yielding  (red trace labeled `Experiment' in the figure). Throughout the testing process, the packings were compressed at a rate of $10$~mm/min.
	
	\begin{table}
      \centering
  	\renewcommand{\arraystretch}{1.5}
  	\renewcommand{\tabcolsep}{0.2cm} 
  	\begin{tabular}{|cl|c|c|}
  		\hline
  		\multicolumn{1}{|l}{} & \multicolumn{1}{l|}{Shape} & $\phi_0$ & $\phi_{cyc}$ \\
  		\hline
  		\includegraphics[width=\littlepic]{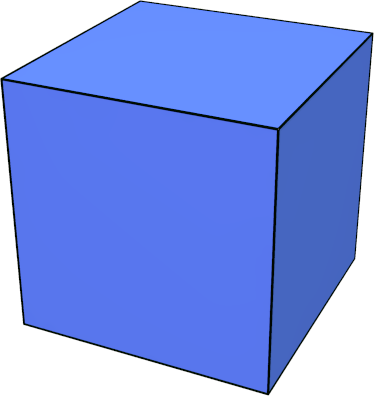} \includegraphics[width=\littlepic]{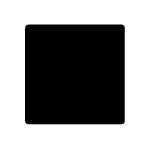}   & Cubes             & 0.59$\pm$0.04 & 0.58$\pm$ 0.04 \\
  		\includegraphics[width=\littlepic]{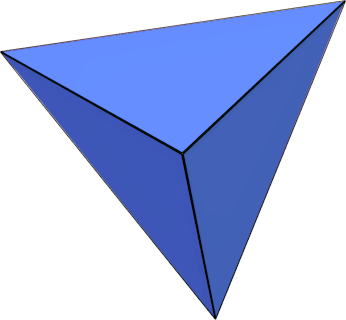} \includegraphics[width=\littlepic]{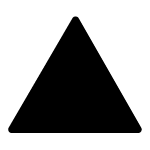}  & Tetrahedra        & 0.57$\pm$0.04 & 0.57$\pm$ 0.04 \\
  		\includegraphics[width=\littlepic]{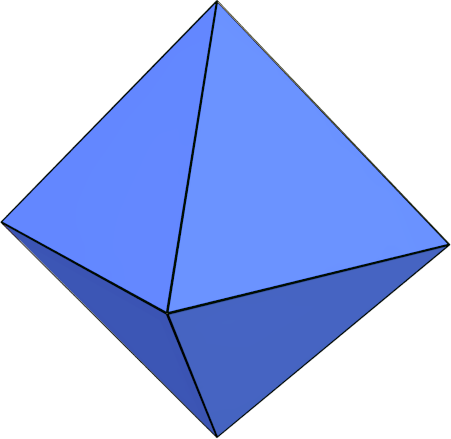} \includegraphics[width=\littlepic]{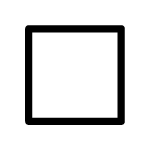}   & Octahedra	        & 0.57$\pm$0.04 & -- \\
  		\includegraphics[width=\littlepic]{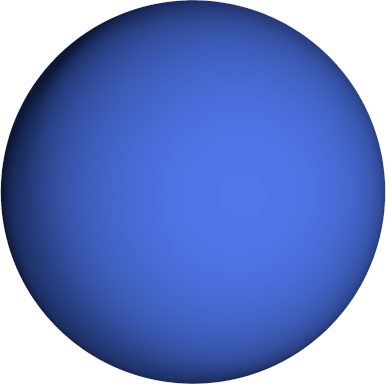} \includegraphics[width=\littlepic]{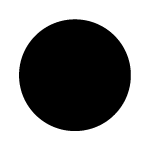} & Spheres 	        & 0.56$\pm$0.04 & 0.55$\pm$ 0.04 \\
  		\includegraphics[width=\littlepic]{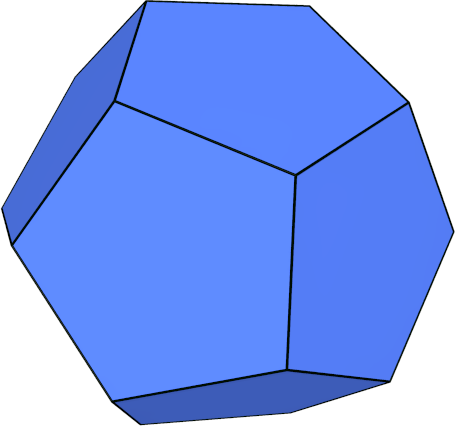} \includegraphics[width=\littlepic]{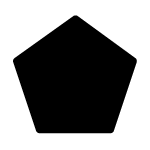} & Dodecahedra	    & 0.56$\pm$0.04 & -- \\
  		\includegraphics[width=\littlepic]{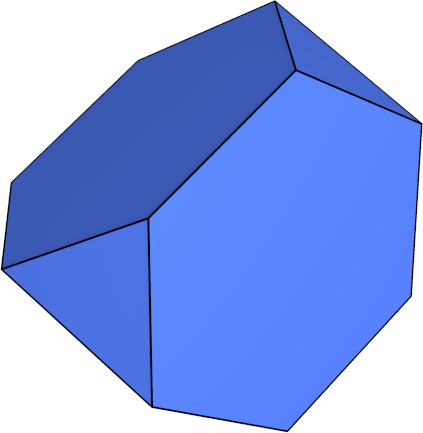} \includegraphics[width=\littlepic]{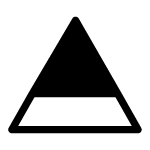} & Trunc. Tetra.	& 0.56$\pm$0.04 & -- \\
  		\includegraphics[width=\littlepic]{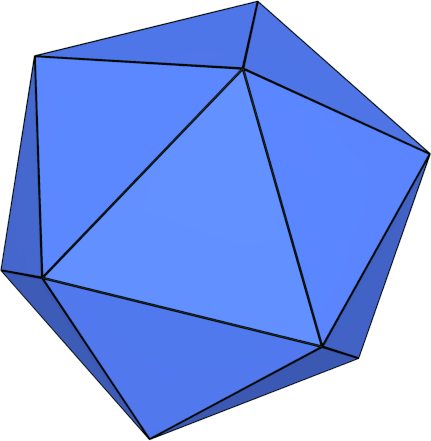} \includegraphics[width=\littlepic]{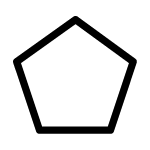} & Icosahedra 	    & 0.55$\pm$0.04 & -- \\
  		\includegraphics[width=\littlepic]{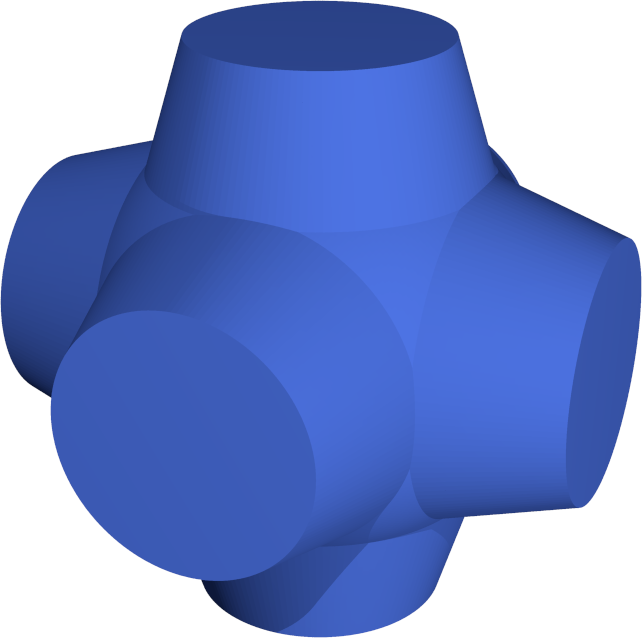} \includegraphics[width=\littlepic]{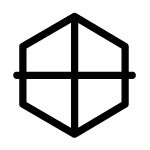} & 0.9mm Jacks	& 0.54$\pm$0.04 & -- \\
  		\includegraphics[width=\littlepic]{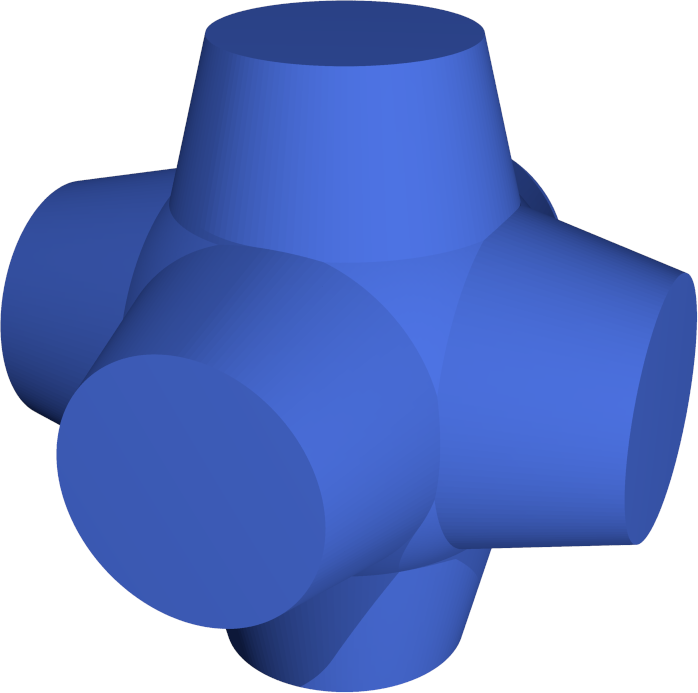} \includegraphics[width=\littlepic]{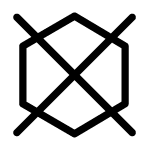} & 1.3mm Jacks		& 0.52$\pm$0.03 & -- \\
  		\includegraphics[width=\littlepic]{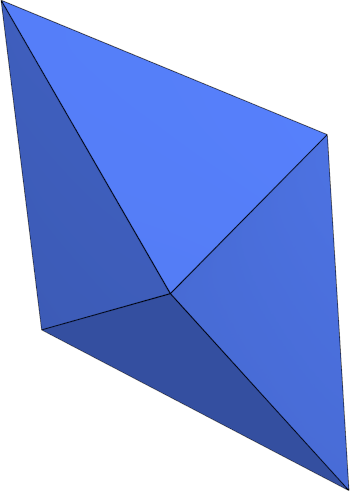} \includegraphics[width=\littlepic]{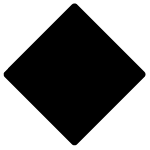} & Tri. Bipyr.	& 0.48$\pm$0.03 & -- \\
 		\includegraphics[width=\littlepic]{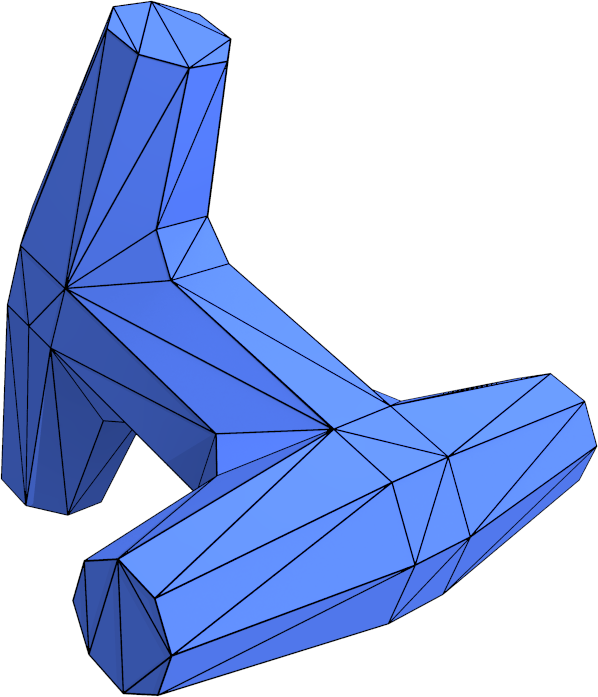} \includegraphics[width=\littlepic]{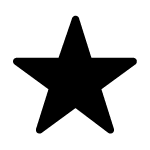} & Dolos 		        & 0.46$\pm$0.03 & -- \\
  		\includegraphics[width=\littlepic]{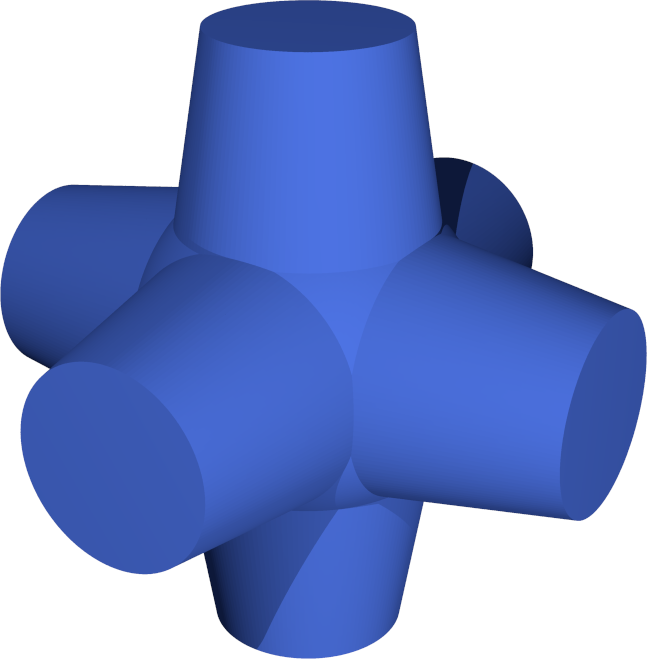} \includegraphics[width=\littlepic]{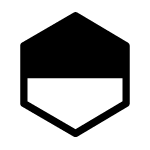} & 2.6mm Jacks 	& 0.46$\pm$0.03 & -- \\ 
  		\includegraphics[width=\littlepic]{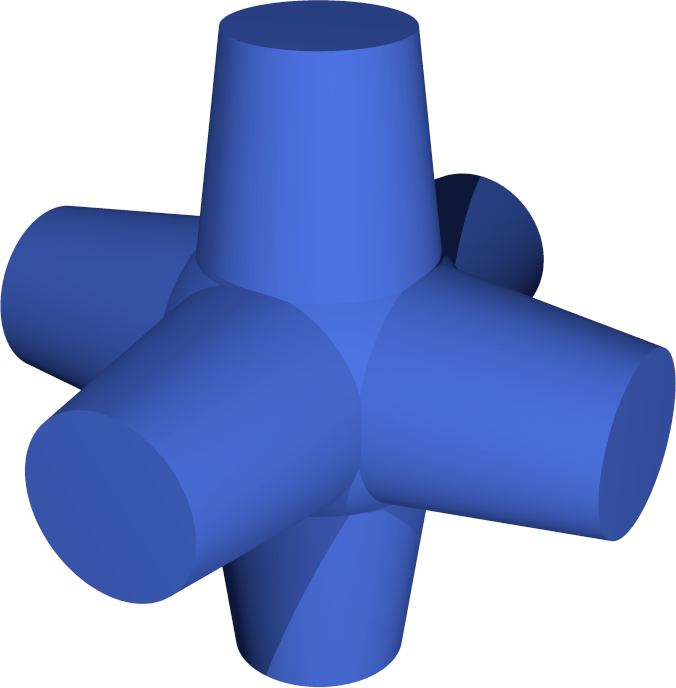} \includegraphics[width=\littlepic]{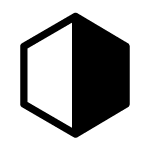} & 3.6mm Jacks     & 0.39$\pm$0.03 & -- \\
  		\includegraphics[width=\littlepic]{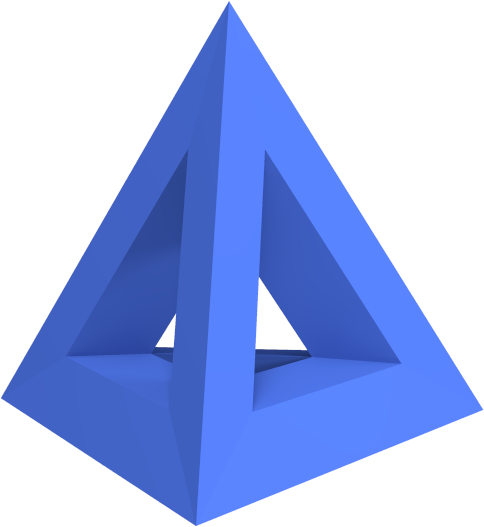} \includegraphics[width=\littlepic]{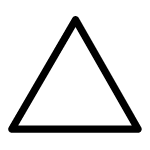} & Tet. Frames	    & 0.25$\pm$0.02 & 0.25$\pm$ 0.02 \\
  		\hline
  	\end{tabular}
  	\caption{Packing fractions $\phi_0$, measured as poured and before applying confinement or conditioning the sample by cyclic loading, and $\phi_{cyc}$, measured for selected shapes after confinement to 0.080~MPa and conditioning by cyclic loading. On the left, next to the rendered shapes, we indicate the corresponding data symbols used throughout this paper.}
      \label{tab:phi}
  	\end{table}

	For each particle shape, we performed 3-5 experimental runs at each confining pressure, starting every test with a freshly poured sample and performing the cyclic conditioning before recording the  stress-strain curves. These repeated runs provided an indication  not only of the reproducibility of the initial conditions but also of shape-dependent fluctuations from run to run. For the results reported below, we analyzed each run individually and extracted response parameters such as the compressive Young's modulus and yield stress. In plots showing ensemble-averaged data, the width of the shaded band surrounding a curve is twice the ensemble standard deviation. The measurement resolution was $8\times10^{-5}$ for strain and  $2\times10^{-4}$~MPa for stress.

	Packing fractions were measured at two times during the experiments. For all shapes, the as-poured fractions (before applying confining stress and  cyclic conditioning) were determined from the known density of the cured plastic, the height and radius of the packing, and the mass of the particles in the sample. Systematic and statistical uncertainties in the measurements resulted in an overall uncertainty of $\approx$6\% in the value of $\phi_0$. For several of the shapes (tetrahedra, cubes, spheres, tetrahedral frames) we performed additional volumetric measurements to track changes in the packing fraction as the sample was being compressed. To perform the volumetric tests, we placed whole sample assembly shown in Fig.~\ref{fig:Setup}a in a sealed, water-filled chamber with a single output line leading to a water bath on a scale (details provided in Suppl. Mat.). The packing's volume changes were monitored by measuring the amount of water in the bath throughout the compression test. The resolution for these volumetric measurements was 5~mm${}^3$ out of 2$\times10^5$~mm${}^3$ sample volume, allowing us to track changes  in $\Delta\phi/\phi_0$ with a resolution of ~1$\times10^{-4}$. Therefore, the uncertainties listed in Table 1 for the packing fraction $\phi_{cyc}$ are dominated by the uncertainties in the starting value $\phi_0$.


	\section*{Results and Discussion}
	\subsection*{Packing Densities}

	Table~\ref{tab:phi} shows the as-poured packing fractions, $\phi_0$, for all particle shapes tested. For some of the shapes comparisons can be made with prior work. In agreement with experiments  on plastic dice by Baker et al.,\cite{BakerPlat} our packings of the Platonic solids follow a non-monotonic trend with increasing number of faces per particle, exhibiting a slight peak in the mean values of $\phi$ at 6 faces (cube).   One important aspect is that the experimentally measured packing fraction will depend on the packing protocol, especially if the particles are frictional, as in our case.  This friction is due to the properties of the polymeric material used in the 3D printing process as well as the fact that the printing introduces roughness on the scale of the print resolution.  We characterized the resulting friction by measuring the angle of maximal stability in tilt experiments, i.e., the angle before the onset of avalanching (see Suppl. Mat.).  The mean angle obtained for our spheres, 26 degrees, is comparable to values obtained in rotating drum experiments for similar size glass beads \cite{Jaeger1988, Robinson2002}.  As a consequence of friction, the as-poured samples in our experiments,  even for spheres, form loose packing configurations with packing fractions far below the densest possible.  For numerical comparison, the $\phi_0$ values in Table~\ref{tab:phi} correspond most closely to results obtained with the sequential deposition protocol of Baker et al.\cite{BakerPlat}  Adding tapping or vibrating to the preparation protocol can increase the density considerably.  For tetrahedra this was recently  investigated  also by Neudecker et al. \cite{NeudeckerTets}

	\begin{table*}[t]
      \centering
  	\renewcommand{\arraystretch}{1.5}
  	\renewcommand{\tabcolsep}{0.3cm} 
  	\begin{tabular}{|cl|c|c|c|c|c|c|c|c|c|}
  		\cline{4-9} 
  		\multicolumn{3}{c|}{} & \multicolumn{3}{c|}{$\sigma_{con}=0.001$~MPa} & \multicolumn{3}{c|}{$\sigma_{con}=0.080$~MPa} \\
  		\hline
  		\multicolumn{1}{|l}{} & \multicolumn{1}{l|}{Shape} &  $\theta_m$ (deg) & $E$ (MPa) & $\sigma_y$ (kPa) & $\varepsilon_y$ ($\times10^{-3}$) & $E$ (MPa) & $\sigma_y$ (kPa) & $\varepsilon_y$ ($\times10^{-3}$)\\
  		\hline
  		\includegraphics[width=\littlepic]{Jack36mm_ren.png} \includegraphics[width=\littlepic]{Jack36mm_sym.png} & 3.6mm Jacks     & 57$\pm$3 & 6.63$\pm$0.15  &  9.51$\pm$0.11  &  1.43$\pm$0.04  &  59.5$\pm$0.2  &  167$\pm$1  &  3.86$\pm$0.02 \\
 		\includegraphics[width=\littlepic]{Dolos_ren.png} \includegraphics[width=\littlepic]{Dolos_sym.png} & Dolos 		        & 41$\pm$3 & 9.67$\pm$0.33  &  5.25$\pm$0.08  &  0.93$\pm$0.03  &  63.2$\pm$0.3  &  109$\pm$1  &  2.54$\pm$0.01 \\
  		\includegraphics[width=\littlepic]{Jack26mm_ren.png} \includegraphics[width=\littlepic]{Jack26mm_sym.png} & 2.6mm Jacks 	& 39$\pm$3 & 7.62$\pm$0.25  &  5.51$\pm$0.13  &  0.97$\pm$0.04  &  68.6$\pm$0.3  &  162$\pm$1  &  3.15$\pm$0.01 \\
  		\includegraphics[width=\littlepic]{Bipyramid_ren.png} \includegraphics[width=\littlepic]{TriBiPyr_sym.png} & Tri. Bipyr.	& 39$\pm$3 & 4.14$\pm$0.20  &  12.65$\pm$0.03  &  3.41$\pm$0.18  &  71.2$\pm$0.3  &  324$\pm$1  &  5.53$\pm$0.03\\
  		\includegraphics[width=\littlepic]{TruncTet_ren.png} \includegraphics[width=\littlepic]{TrTet_sym.png} & Trunc. Tetra.	& 39$\pm$3 & 3.15$\pm$0.10  &  8.19$\pm$0.07  &  2.59$\pm$0.09  &  77.9$\pm$0.3  &  292$\pm$1  &  4.41$\pm$0.02 \\
  		\includegraphics[width=\littlepic]{Tetra_ren.png} \includegraphics[width=\littlepic]{Tetra_sym.png} & Tetrahedra        & 37$\pm$3 & 5.60$\pm$0.16  &  7.83$\pm$0.16  &  1.72$\pm$0.05  &  82.0$\pm$0.3  &  380$\pm$1  &  5.38$\pm$0.02 \\
  		\includegraphics[width=\littlepic]{Octa_ren.png} \includegraphics[width=\littlepic]{Octa_sym.png}   & Octahedra	        & 37$\pm$3 & 3.16$\pm$0.29  &  8.84$\pm$0.27  &  3.15$\pm$0.30  &  90.0$\pm$0.3  &  340$\pm$1  &  4.47$\pm$0.02 \\
  		\includegraphics[width=\littlepic]{TetFrame_ren.png} \includegraphics[width=\littlepic]{TetFrame_sym.png} & Tet. Frames	    & 37$\pm$3 & 2.47$\pm$0.14  &  8.75$\pm$0.04  &  3.68$\pm$0.24  &  33.3$\pm$0.2  &  171$\pm$1  &  6.45$\pm$0.05 \\
  		\includegraphics[width=\littlepic]{Jack13mm_ren.png} \includegraphics[width=\littlepic]{Jack13mm_sym.png} & 1.3mm Jacks		& 35$\pm$3 & 6.05$\pm$0.17  &  4.96$\pm$0.06  &  1.11$\pm$0.03  &  91.3$\pm$0.3  &  203$\pm$1  &  2.99$\pm$0.01 \\
  		\includegraphics[width=\littlepic]{Cube_ren.png} \includegraphics[width=\littlepic]{Cube_sym.png}  & Cubes             & 35$\pm$3 & 3.04$\pm$0.10  &  5.58$\pm$0.14  &  2.46$\pm$0.08  &  77.4$\pm$0.3  &  221$\pm$1  &  3.54$\pm$0.02 \\
  		\includegraphics[width=\littlepic]{Jack092mm_ren.png}  \includegraphics[width=\littlepic]{Jack09mm_sym.png} & 0.9mm Jacks	& 34$\pm$3 & 4.27$\pm$0.16  &  3.45$\pm$0.17  &  1.22$\pm$0.05  &  90.2$\pm$0.3  &  119$\pm$1  &  1.91$\pm$0.01 \\
  		\includegraphics[width=\littlepic]{Icosa_ren.png} \includegraphics[width=\littlepic]{Icosa_sym.png} & Icosahedra 	    & 34$\pm$3 & 2.80$\pm$0.14  &  4.49$\pm$0.09  &  2.45$\pm$0.13  &  74.7$\pm$0.4  &  210$\pm$1  &  3.45$\pm$0.02 \\
  		\includegraphics[width=\littlepic]{Dodeca_ren.png} \includegraphics[width=\littlepic]{Dodeca_sym.png} & Dodecahedra	    & 34$\pm$3 & 2.74$\pm$0.10  &  8.30$\pm$0.06  &  3.61$\pm$0.14  &  73.5$\pm$0.3  &  268$\pm$1  &  4.43$\pm$0.02 \\
  		\includegraphics[width=\littlepic]{Sphere_ren.png} \includegraphics[width=\littlepic]{Sphere_sym.png} & Spheres 	        & 26$\pm$3 & 3.24$\pm$0.19  &  1.84$\pm$0.12  &  0.92$\pm$0.06  &  60.2$\pm$0.2  &  147$\pm$1  &  3.04$\pm$0.01 \\
  		\hline
  	\end{tabular}
  	\caption{Characteristic properties for each shape. The angle of maximum stability ($\theta_m$), and mechanical properties  (effective modulus $E$, yield stress $\sigma_y$, yield strain $\epsilon_y$) of random packings at the lowest and highest confining pressures ($\sigma_{con}$) tested. Values shown are the weighted averages of all experimental runs under the same conditions. See text and Fig. 6 for operational definitions of $\sigma_y$ and $\epsilon_y$. }
      \label{tab:ang}
  	\end{table*}

	Among the convex shapes that are not part of the family of Platonic solids, (frictionless) truncated tetrahedra have been found in computer simulations\cite{DamaTet} to pack particular densely.  As Table~\ref{tab:phi} shows, this does not translate to large $\phi_0$ for poured, random packings of their frictional counterparts.  The triangular bipyramid clearly stands out with a remarkably low $\phi_0$ = 0.48. Its elongated shape creates configurations that, in terms of porosity (1- $\phi_0$), start to compete with non-convex shapes such as jacks.  As the arm length of the jacks is increased, we find that  $\phi_0$ decreases significantly, from values close to that for spheres down to 0.39. Our most porous packings were those comprised of  tetrahedral frames, owing primarily to their hollow interior.
	
	As described earlier, our sample conditioning protocol involved the application of a confining pressure followed by cyclic axial loading/unloading. We define $\phi_{cyc}$ as the packing fraction in the unloaded ($q$=0) state at the end of N = 10 loading/unloading cycles. Since the very first loading cycle already applied a vertical displacement of 3\% of the sample height, this conditioning produces dilation. Thus, the unloaded packing configuration $\phi_{cyc}$ can at best recover to $\phi_0$ and will typically remain somewhat smaller.  As Table~\ref{tab:phi} shows for several selected shapes, we find packing fractions $\phi_{cyc}$ about 0.01 lower than $\phi_0$.

	\subsection*{Mechanical Response}

\begin{figure}
        \centering
		\includegraphics[width=.90\narrowf]{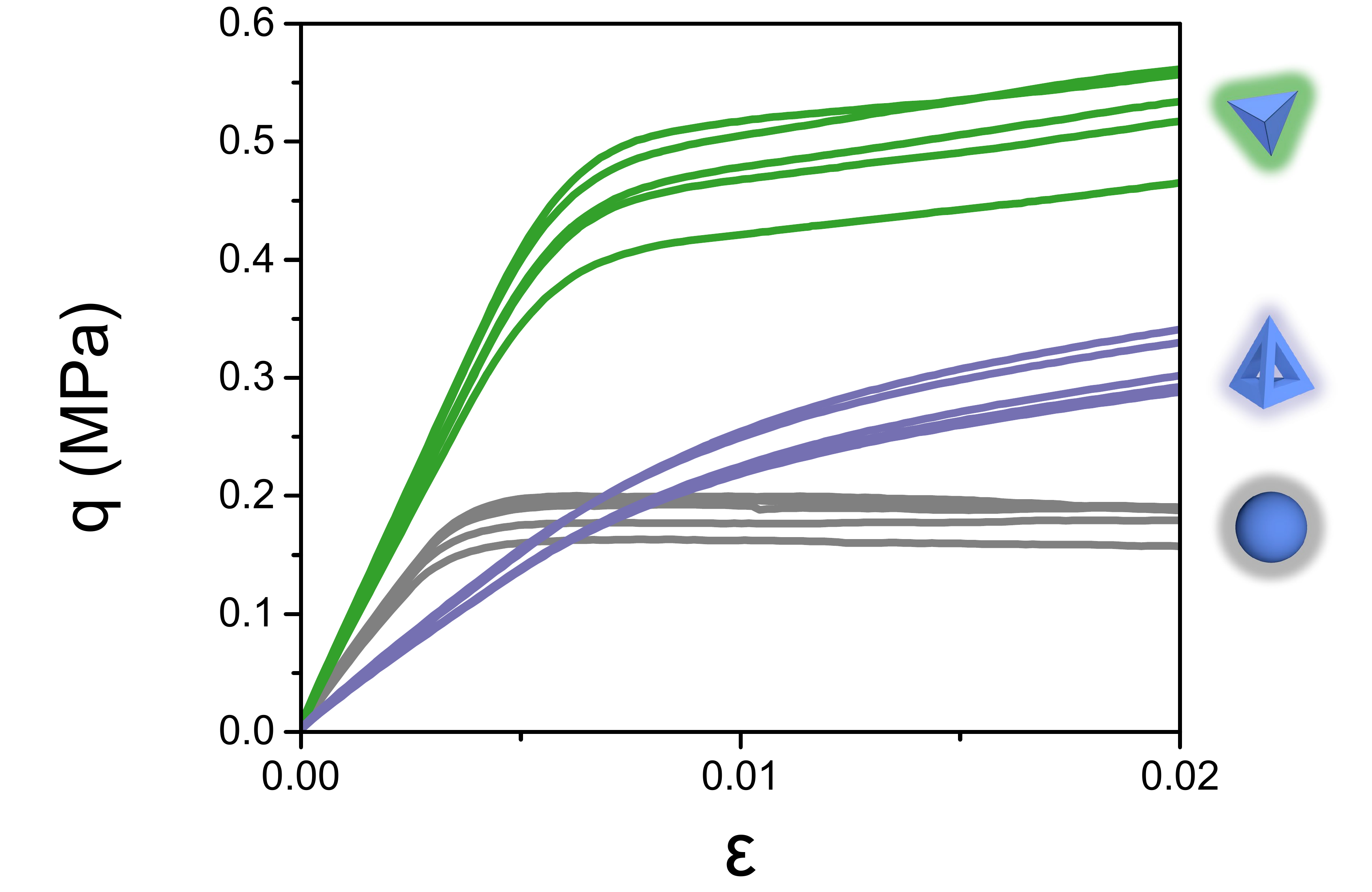}
		\caption{Raw stress-strain curves at $\sigma_{con}$=0.080~MPa for spheres, tetrahedra, and tetrahedral frames. Each curve represents a separate experiment.}
        \label{fig:SSRaw}
        
		\includegraphics[width=.90\narrowf]{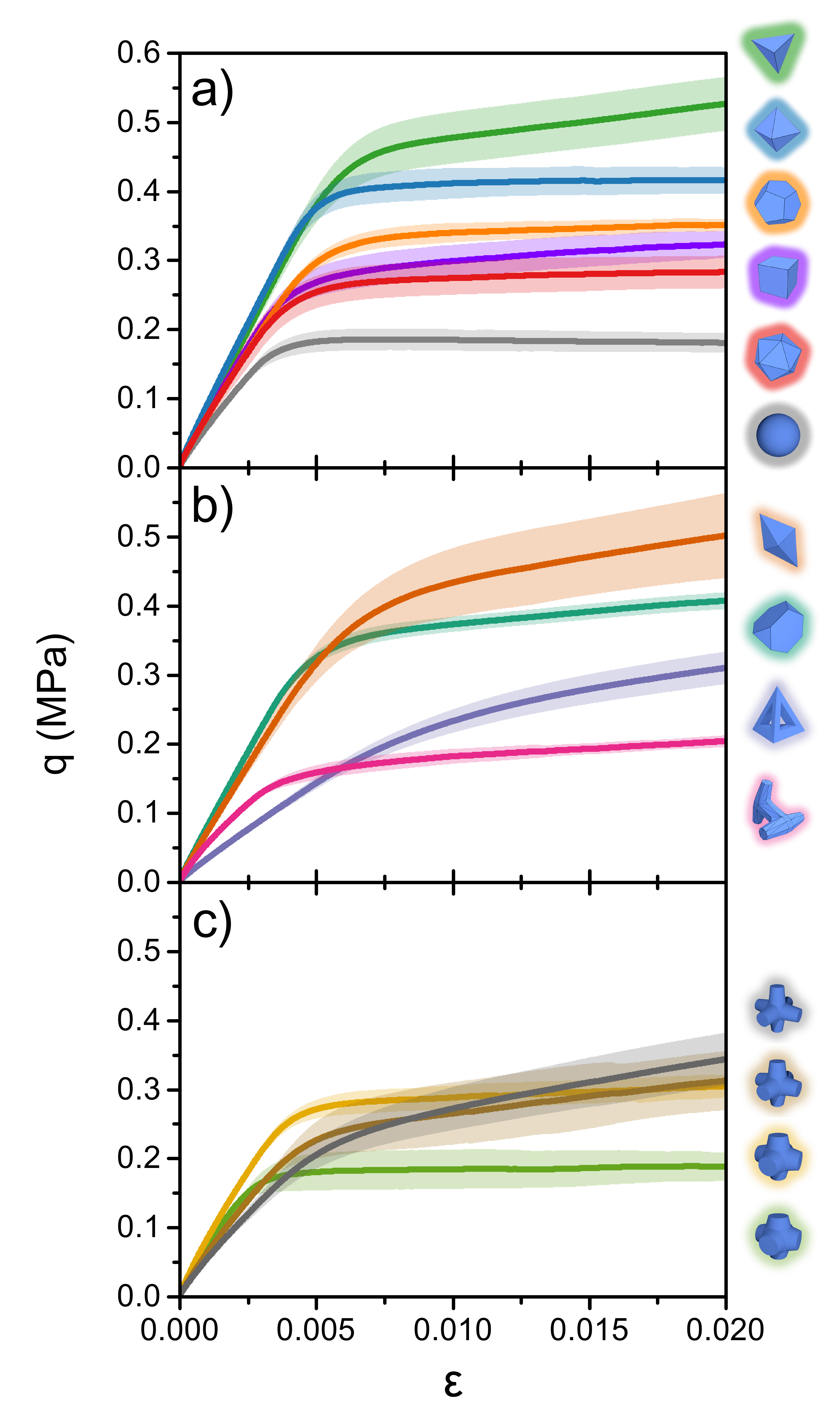}
		\caption{Ensemble-averaged stress-strain curves at  $\sigma_{con}=0.080$~MPa. Solid lines represent averages of 3-5 independent tests for each shape. The half width of the shaded bands represents one standard deviation. The vertical order of the shapes drawn along the side, as well as the color of their outlines, corresponds to the large strain ordering of the curves (and their color). }
        \label{fig:SSAvg} 
	\end{figure}

    For all shapes tested, the stress-strain curves exhibited two regimes. At small strain the stress increased in approximately linear fashion. For larger strain, the stress smoothly transitioned into a plastic failure regime, characterized by a significant reduction in the slope.  Due to the large number of particles in each packing as well as the sample conditioning, individual experimental runs produced very clean and smooth traces (significant fluctuations in individual traces appeared only in a regime corresponding to large scale sample deformation at much higher strains, not discussed here).  In Fig.~\ref{fig:SSRaw} we plot examples of stress-strain curves from individual runs to give an idea about the (shape-dependent) run-to-run variability.
	  
	Figure~\ref{fig:SSAvg}a shows the ensemble averaged traces for each of the five Platonic solids and spheres at a confining pressure of 0.080~MPa. The equivalent plots for the other shapes tested are shown in Fig.~\ref{fig:SSAvg}b,c. Inspection of these plots reveals a number of trends. Compared to spheres (bottom trace in Fig.~\ref{fig:SSAvg}a) the introduction of facets in the Platonic solids increases the initial slope, i.e., the packing's stiffness, as well as the stress level at which large-scale plastic deformation sets in. Relatively compact, sphere-like shapes (including highly faceted particles as well as jacks with very short arms) exhibit nearly perfectly plastic failure beyond yielding, i.e., a nearly horizontal  $q(\varepsilon)$. Shapes with sharp points (tetrahedra) or significant protrusions (long-armed jacks), on the other hand,  even after yielding produce significant further increases in stress. 

	One immediate observation from this data is that high packing density does not directly correlate with particularly stiff and strong packings. For example, while cubes pack most densely under our sample preparation conditions, they exhibit a smaller initial slope and lower onset stress for yielding than the tetrahedra and octahedra (Fig.~\ref{fig:SSAvg}a), which pack at lower density  (Table~\ref{tab:phi}).   The sequence of jacks (Fig.~\ref{fig:SSAvg}c) demonstrates this point explicitly and highlights the non-monotonic dependence on changes in particle geometry:   At 0.92mm arm length the packing's response is essentially identical to that of spheres, i.e., the additional `geometric friction' from the short protrusions hardly matters.  At 1.3mm arm length $\phi_0$ has decreased significantly, but now the interpenetrating arms enable a significant enhancement in both stiffness and strength, while still keeping the overall shape of the stress-strain curve similar, including the nearly perfect plastic failure regime.  Doubling the arm length to 2.6mm decreases the initial stiffness, presumably because of the concomitant 6\% reduction in packing density, but now introduces a significant residual stiffness in the plastic failure regime.  Finally, with 3.6mm arm length, the jacks pack so loosely that the packing's initial load response becomes quite soft. However, at larger strains the interpenetrating arms enable rapidly increasing levels of stress, to the point that, within the range up to $\epsilon$ = 0.02 plotted,  the stress supported by the packing exceeds that of all the other jacks. 

\begin{figure*}
      \centering
      \includegraphics[width=.87\widef]{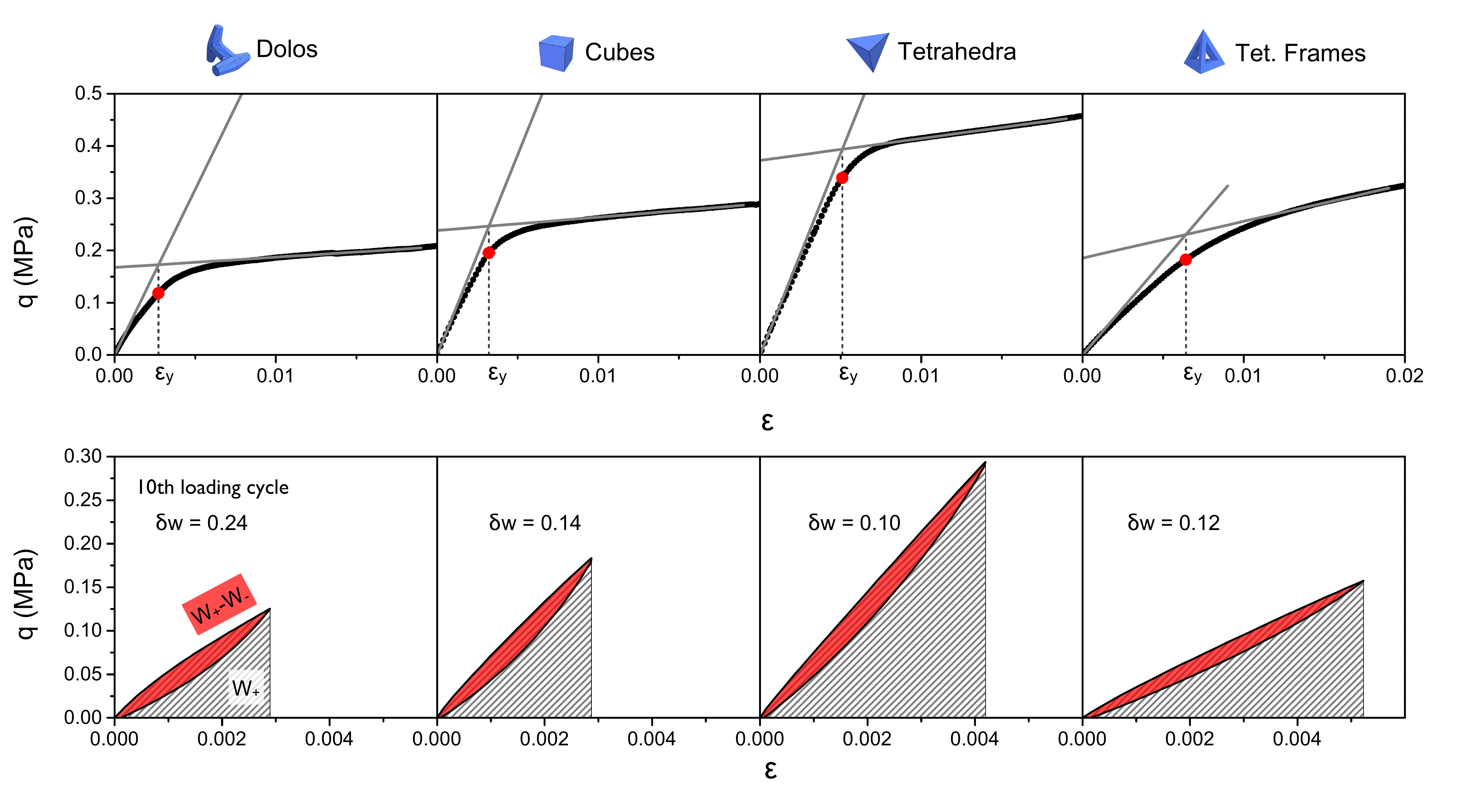}
    	\caption{Analysis of mechanical response. Data are for selected shapes at $\sigma_{con}$ = 0.080~MPa. Top row: Dark lines are experimental stress-strain data,  gray lines represent  linear fits for the low- and high-strain  regimes.  The intersection of the gray lines operationally defines a yield strain $\varepsilon_y$ and a yield stress $\sigma_y = \sigma(\varepsilon_y)$, shown as red dot. Bottom row: Energy loss during the final conditioning cycle. The net loss, highlighted in red, is the difference in work performed  during loading ($W_{+}$, striped region), and unloading ($W_{-}$).  The relative loss is indicated by $\delta w$.}
    \label{fig:Calc}
	\end{figure*}

	In order to compare the performance of various shapes more systematically, we use the stress-strain curves to calculate an effective Young's modulus under compression, a yield stress, and the energy lost during cycling (as shown in Fig.~\ref{fig:Calc}). The modulus $E = \lim_{\varepsilon\to0}dq/d\varepsilon$  corresponds to the slope of the stress-strain curves in the small-$\varepsilon$ limit, i.e., the initial stiffness. Since we do not know the functional form of the shape-dependent load response $q(\varepsilon)$, we expand around $\varepsilon=0$, fit to a quadratic form $$q(\varepsilon)=E\varepsilon + \frac{1}{2}\frac{d^2q}{d\varepsilon^2}\varepsilon^2 $$ ignoring $O(\varepsilon^3)$ terms, and take the fit's linear coefficient as the modulus. We find that including the second-order terms is important to obtain meaningful and reproducible results. Because we do not a priori know the extent of the region that can be approximate by a linear stress-strain relationship, we varied the strain range included in each fit, selecting the fit with the best $\chi^2$ value. This method of fitting tended to include data data up to strain levels $\varepsilon=0.002$.

	The yield stress is a measure of the strength of a given packing. In granular materials, it can be defined in multiple ways, depending on whether the focus is on the maximum stress sustained or on the stress level at which deviations from linear behavior first set in. The angle of maximum stability of a heap of granular material provides a measure of the yield stress under shear and conditions of very small confinement (self-weight of the particles at the free surface). In Table 2 we list this angle for all shapes tested (see Appendix for measurement details).  For evaluating the strength under compression, we operationally associate the yield stress with  the stress level at which $q(\varepsilon)$ crosses over to the second regime, in which it undergoes significant plastic deformation. To quantify the cross-over point we linearize the initial regime as done with the modulus calculation and the second regime by fitting a line to the region $\varepsilon\in(0.12,0.18)$. We associate the intersection of these lines with the packing's yield strain $\varepsilon_y$ and take the yield stress to be $\sigma_y=q(\varepsilon_y)$. This is illustrated in Fig.~\ref{fig:Calc}, top row, for several shapes.

	The third parameter we use to characterize the behavior of different particle shapes is the energy per volume dissipated during conditioning. We extract this from the area enclosed by cyclic loading/unloading loops recorded while conditioning the sample. To compare the degree of energy loss among different shapes, we use the final loading/unloading cycle to calculate $\delta w = (W_{+} - W_{-})/W_{+}$, the relative difference in mechanical work performed during loading ($W_{+}$) and unloading ($W_{-}$) (Fig.~\ref{fig:Calc}, bottom row). 

\begin{figure}[!t]
        \centering
        \includegraphics[width=\narrowf]{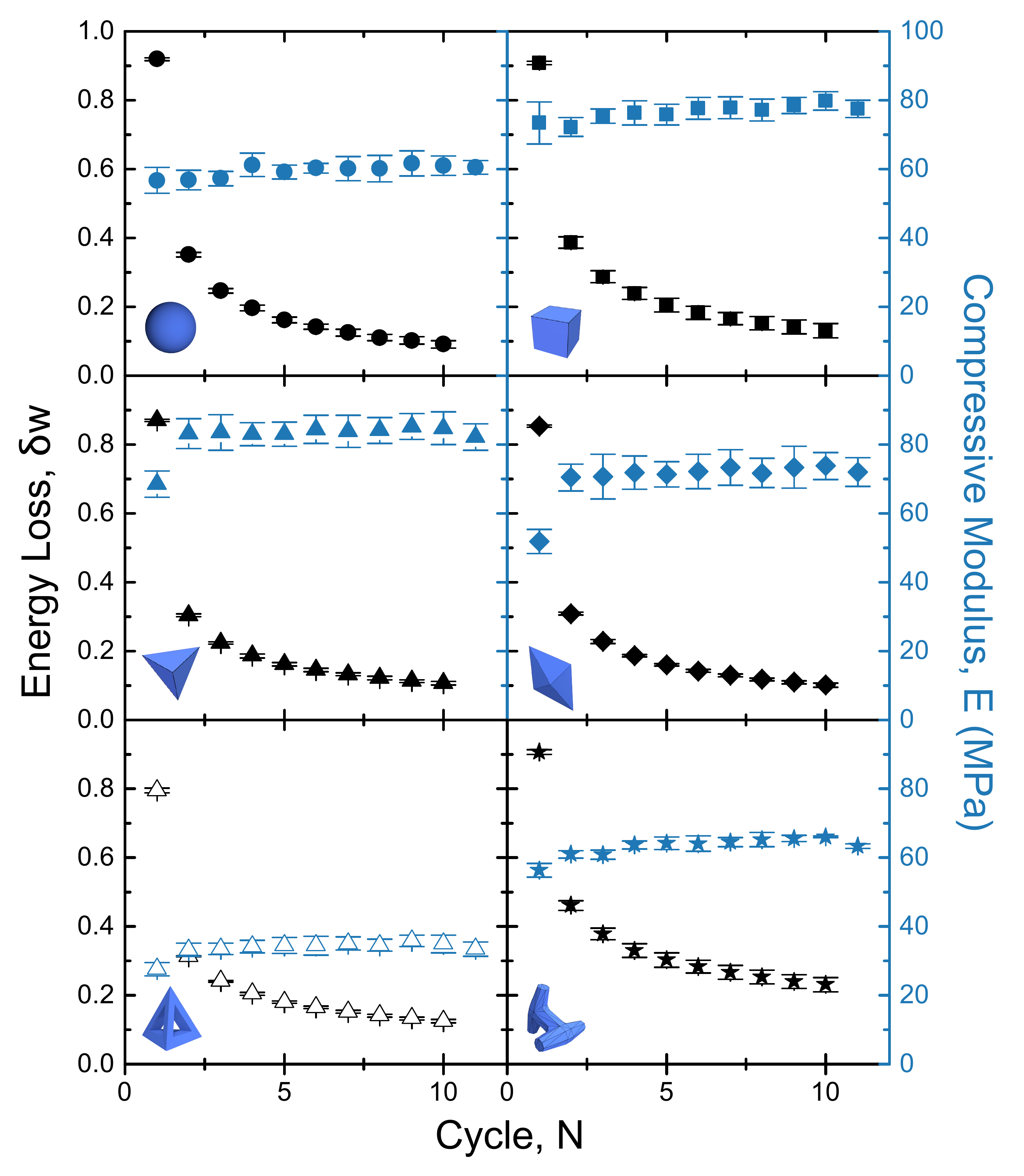}
        \caption{Evolution of the mechanical response during cyclic loading as part of the conditioning of the aggregates. Data for effective modulus, $E$, and energy loss per cycle, $\delta w$, are plotted as function of number of cycles, $N$, for representative particle shapes at 0.080~MPa confinement. }
     	\label{fig:Cycles} 
    \end{figure}

	The evolution of modulus $E$ and energy loss per cycle $\delta w$ with $N$ is shown in Fig.~\ref{fig:Cycles}. 
	Statistical fluctuations for each shape are indicated by the error bars. The values for $E$ shown at $N$ = 11 are the average effective moduli after conditioning. We note that the stiffness of the response to small loading, parameterized by $E$, quickly settles into an asymptotic value after a few cycles. This occurs independent of particle geometry and whether the shape is convex or not.  By contrast, the energy loss per cycle changes more slowly and keeps decreasing even after $E$ has leveled off. We emphasize  that $\delta w$ is the relative energy loss, defined as fraction of the (also decreasing) energy input per cycle.  As such, it provides a measure of the non-elastic deformation associated with structural rearrangements in the packing during each loading/unloading cycle. Since the strain applied during loading decreases with $N$ these rearrangements become smaller too, but  they will continue at least as long as each cycle exceeds the yield strain, i.e., exceeds the regime over which the response is effectively linear with modulus $E$.

    \subsection*{Mapping the Relationship between Materials Parameters, Shape, and Confinement}
    
    So far, we discussed the mechanical response at confinement pressure $\sigma_{con}$ = 0.080~MPa. Repeating the same tests over a range of  $\sigma_{con}$, and each time extracting the material parameters $E$,  $\sigma_y$ and $\delta w$ as described above, we can build up a more comprehensive mapping of the stress response.  Using confining pressure as an independent variable also allows us  to correlate the material parameters in Ashby-type plots.\cite{WegstEfficiency} Such plots demonstrate where each shape falls in the response space, thereby inviting not only comparisons between each other, but also comparisons with other materials. 
        
\begin{figure*}
        \centering
		\includegraphics[width=.9\widef]{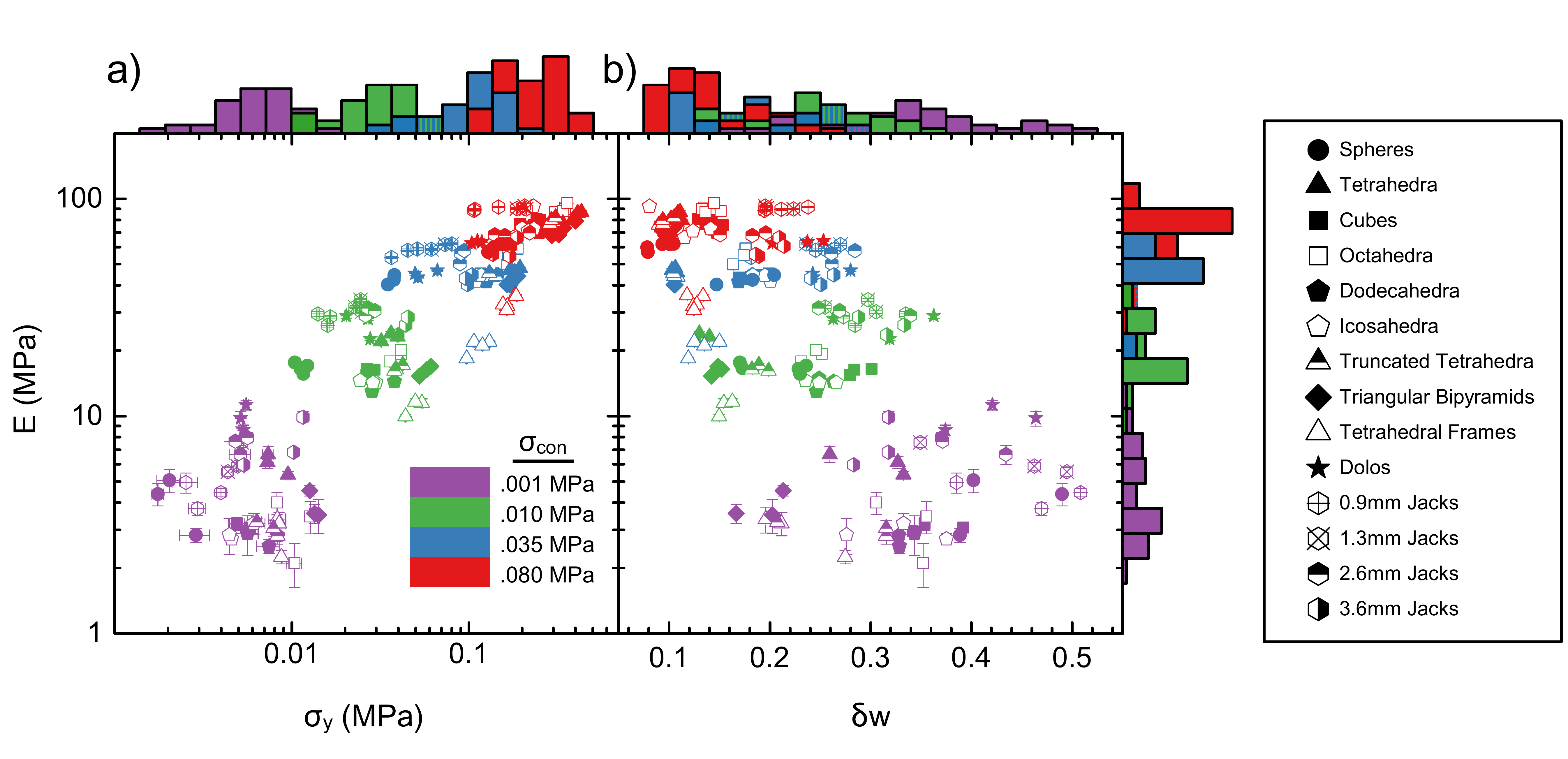}
		\caption{Relationships among the effective material parameters stiffness, strength, and energy loss per cycle. Data for all packings tested are shown. \textbf{(a)} Relationship between $E$ and $\sigma_y$.  \textbf{(b)}  Relationship between $E$ and $\delta w$. A key to the data symbols for different particle shapes is provided on the right. Colors correspond to the confining pressures as listed in the inset to panel (a). Histograms along the edges of the plots are projections of the data into a binned point density  along the axis. They provide an indication of the range across all shapes of the material parameters at particular confinement pressures. If two bars of different color completely overlap, the bar is shown with pinstripes of both colors.}
        \label{fig:Behavior}
	\end{figure*}

    For every triaxial test we performed (3-5 per particle type at each pressure), Fig.~\ref{fig:Behavior}a shows  packing stiffness (Young's modulus $E$) plotted against packing strength (yield stress $\sigma_y$). Fig.~\ref{fig:Behavior}b is the same type of plot but relating packing stiffness to energy loss ($\delta w$).  Different symbols represent different shapes, while colors indicate the confining pressure. Comparing symbols of the same shape and color gives an indication of the variation in material parameters among an ensemble of identically prepared samples.  The histograms along the axes help to identify how, for each confining pressure, the different parameters are distributed among all shapes tested. 
    
   	Several features jump out from Fig.~\ref{fig:Behavior}a. Firstly, the data as a whole occupy a well-defined region along the diagonal, indicating that $E$ and $\sigma_y$ are correlated. Secondly, with varying confining stress the data for individual shapes move along the diagonal across several orders of magnitude in $E$ and $\sigma_y$. For given confinement $\sigma_{con}$, in turn, shape plays a more nuanced role by tuning the aggregate response around the shape-averaged behavior.
    
    The strong dependence on $\sigma_{con}$ is a particularly special characteristic of granular material as a class. Because the material's strength under compression is linked to the confining pressure, it can attain a wider range of values than most other materials, including natural materials, which typically are limited to about one order of magnitude in modulus and/or strength.\cite{WegstEfficiency} In fact, our one and a half decades in $E$ are only a subset of the possible range. Going lower, we expect that the response could extend another order of magnitude before gravitational stress scales begin to come into play. Going higher, for example by pressurizing the chamber containing the sample, the limit will be set by the performance of the particle material.  For our 3D-printed particles, we expect that the response might extend another half order of magnitude (in  Fig.~\ref{fig:Behavior}a the effective moduli of the various packings appear to begin leveling off around 100 MPa, roughly 1/10 the modulus of the constituent plastic). 
    
    For given confinement, we find that particle shape provides a control knob that can tune the yield stress $\sigma_y$ by about one order of magnitude. This range is largely independent of $\sigma_{con}$, and shapes that produce a low yield stress, such as spheres, consistently tend to be at the tail end of the set of shapes investigated, while others, such as the bipyramids, maintain a high $\sigma_y$ throughout.  For the modulus $E$, on the other  hand, the range becomes much broader  as $\sigma_{con}$ is lowered, from spanning values that differ by a factor of 2 at 0.080~MPa to nearly  a decade at 0.001~MPa. In other words, the role of particle shape becomes significantly more pronounced at low confining pressure.  This can be seen qualitatively from the histograms along the edges of  Fig.~\ref{fig:Behavior} and we will return to it in more quantitative detail later.
        	
    Close inspection of Fig.~\ref{fig:Behavior}a shows that some shapes trade places in their performance when $\sigma_{con}$ is varied.  This highlights how boundary conditions can affect behavior.  For example, dolos and 3.6mm arm length hexapods exhibit the highest modulus among all shapes at 0.001MPa confinement, appropriate considering their use in breakwaters.  But already at 0.01MPa confinement the shorter-armed jacks catch up, taking over as the top performers  in terms of $E$ with increasing $\sigma_{con}$,  and at 0.080~MPa relegating dolos and long-armed jacks to the bottom of the set, next to spheres. Similarly, the Platonic solids (except tetrahedra) produce packings that exhibit the smallest moduli at low confining pressure but cross over to deliver some of the stiffest load responses at 0.080~MPa (in particular the octahedra).  
    
    For convenience and to allow for direct quantitative comparison, an ensemble-averaged subset of the data in Fig.~\ref{fig:Behavior}a is listed in Table 2.  In this table, shapes are listed in order of their maximum angle of stability, $\theta_m$, from avalanche experiments. Clearly, under compression and with increasing confinement the ranking changes.  Furthermore, while long-armed jacks and dolos are particular effective in resisting rolling or sliding under shear and their packings have a large $\theta_m$, they are not especially strong in terms of yielding under compression, where they are outperformed by shapes such as triangular bipyramids or, in the case of dolos, several of the Platonic solids.
    
    It is intriguing that there is no clear separation in performance in a plot of $E$ vs. $\sigma_{y}$ according to whether a particle shape is convex or not, i.e., whether it enables interlocking.   However, with increasing confinement the one particle type tested that allows for interpenetration, the tetrahedral frames, stands out:  while similar to the non-tetrahedral Platonic solids at the lowest $\sigma_{con}$, with increasing $\sigma_{con}$ the frames do not benefit from from stronger interparticle contacts and remain significantly softer (lower in $E$) than all other shapes.
     	
    The fact that inelastic effects are crucial in describing the mechanical response can be seen in the neighboring plot of $E$ vs. $\delta w$ (Fig.~\ref{fig:Behavior}b).  The horizontal axis compares the energy dissipated by a loading/unloading cycle to the energy input.  Immediately it is apparent that a majority of our experiments occur in regions where a significant fraction of the input energy is irrecoverably lost.  For all shapes the relative importance of dissipation decreases at higher confining pressures.   In other words, packings become less inelastic with stronger confinement.  In addition, a number of  shape-dependent trends emerge.  For example, with the exception of a single run at 0.001~MPa, all of the jacks and dolos lie on the right side of the plot for all confining pressures, i.e., consistently exhibit the largest energy loss per cycle. Conversely, for given $E$ the tetrahedral frames exhibit  $\delta w$ values that are among the smallest. Among some of the convex particles the performance changes significantly with $\sigma_{con}$. In particular, truncated tetrahedra and spheres quickly reduce  $\delta w$ as $\sigma_{con}$ increases, with spheres becoming the shapes with the lowest  $\delta w$ at $\sigma_{con}$ = 0.080~MPa.

	\subsection*{Effective Modulus as a Function of Confining Pressure}
 
\begin{figure}[!b]
        \centering
		\includegraphics[width=.9\narrowf]{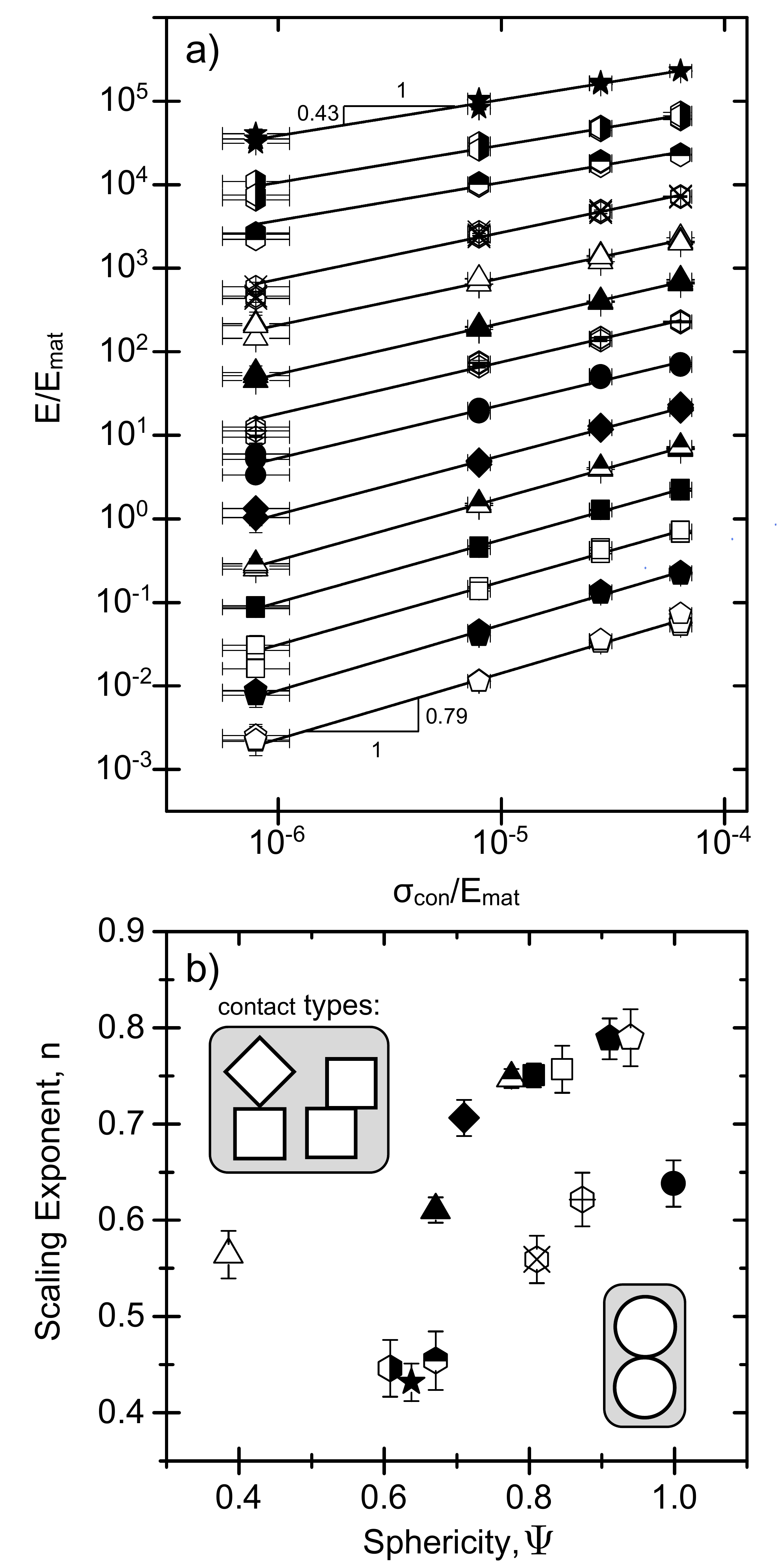}
		\caption{Scaling behavior of effective modulus, $E$, with confinement pressure, $\sigma_{con}$. Symbols are the same as in Fig. 8. \textbf{(a)} Log-log plot indicating that all data are well represented by power laws $E/E_{mat} \propto (\sigma_{con}/E_{mat})^n$, where the exponent characterizes the shape and $E_{mat}$ is the compressive modulus of the 3D-printed plastic. In ordering the modulus data in terms of increasing exponent $n$ (top to bottom), data for each shape except for icosahedra data (bottom trace) were shifted along the y-axis by varying amounts. \textbf{(b)} Scaling exponent $n$ versus particle sphericity $\Psi$. The scaling exponents can be separated into two groups which each follow the same trend of increasing $n$ with $\Psi$.  The upper group includes all polyhedral shapes, which can interact via several types of contacts, depending on relative orientation (see inset). The lower group contains shapes with arms, all of which interact via one type of point-like contact, and includes the spheres as limiting, zero arm-length case.}
        \label{fig:ModScale}
	\end{figure}

    Figure~\ref{fig:ModScale}a shows the dependence of the effective, compressive Young's modulus on confining pressure. Over the range of $\sigma_{con}$ accessible to our experiments, we find that the data are well described by a power law of the form $E \propto (\sigma_{con})^n$.  Physically, the scaling exponent $n$ characterizes how sensitively the packing stiffness reacts to changes in confinement.   The fact that the same functional form  captures the behavior for completely different particle geometries makes $n$ a suitable parameter to investigate how shape affects this sensitivity. 
   
    While values for $n$ reported from experiments on various types of sands are typically close to 0.5, \cite{WoodSoilMech} we observe pronounced shape-dependent differences in $n$ that cover the range from 0.4 to 0.8.  For spheres we can compare the exponent directly with predictions. From Figure~\ref{fig:ModScale}a we have $n \approx 0.64$, significantly larger than the Hertz-Mindlin effective medium theory, which gives $n=1/3$, but consistent with calculations for rough spheres by Yimsiri and Soga.\cite{YimsiriMicromech} 
   
	In order to explore in more detail how $n$ varies  with particle geometry, we parameterize the various shapes by their sphericity\cite{deGraafIrreg,ZouPackChar}, $\Psi=\pi^{1/3} (6V_p)^{2/3} / A_p$.  Here $V_p$ is the particle volume and $A_p$ is its surface area. Particles with larger $\Psi$ are more compact, with $\Psi=1$ corresponding to a perfect sphere. Conversely, small $\Psi$ values indicate a highly non-spherical  shape. 
 
	As Figure~\ref{fig:ModScale}b demonstrates there is a general trend of increasing $n$ with $\Psi$. It is intriguing that highly faceted shapes, and in fact all tested polyhedra with 6 or more faces, have a significantly larger $n$, i.e., a larger sensitivity to $\sigma_{con}$, than spheres.  Furthermore, within the plot of $n(\Psi)$ there seem to be two branches, each exhibiting the same trend of increasing $n$ with $\Psi$:  the lower branch includes the shapes with arms (jacks and dolos) and  at $\Psi$=1 has the spheres as the limiting case of zero arm length, while the upper branch includes the polyhedral shapes. Evidently, higher sensitivity to changes in confinement certainly does not correlate with looser, more porous packing configurations.  In the lower branch in Fig.~\ref{fig:ModScale}b $n$ actually decreases with increasing porosity and in the upper branch  the ordering of the various shapes does not follow their $\phi_0$ values. 
 
	In general, for the modulus to acquire a dependence on confinement pressure, there must be some kind of nonlinearity.  Two obvious sources are nonlinear local contact laws and changes in the packing structure. Because the particles' plastic material should always behave elastically under the  pressures applied, any nonlinearity from particle-particle contact laws must arise from increases in contact area.  If this effect were dominant, it would be natural to see grouping in the exponents $n$ based on the types of contacts allowed for a given shape.  Indeed, Fig. 9 shows that all the particles along the lower branch involve predominately point-like contacts (arm-arm or sphere-sphere).  Conversely, faceted particles that admit more complex interactions (e.g. face-face, face-edge, edge-edge) group together to form the upper branch.  

	Given this observation, a simple picture would be that the surface area increases between preexisting contacts and this in turn generates a non-trivial connection between stiffness and pressure.  However this picture cannot be correct.  In particular, for spheres compression between preexisting contacts would produce a Hertzian contact interaction.  This would yield a packing stiffness $E$ that scales with pressure as $\sigma_{con}^{1/3}$, in contrast to the measured exponent of $n\approx0.64$.

	A more sophisticated scenario might account for the fact that new contacts can be formed as particles are compressed.  For instance, Yimsiri and Soga\cite{YimsiriMicromech} introduced this idea in the form of surface asperities to produce scaling exponents appreciably closer to the observed 0.64.  We note, however, that  for our particles  asperities at the scale of the print resolution (50$\mu$m) would be far too small to significantly alter  the contact law.\cite{YimsiriMicromech} Still, the essential physics behind asperity models is that new contacts are generated, and in disordered packings of frictional objects this could also arise from small displacements between particles that are almost in contact.   
 
	As further evidence that changing contact area and contact number must be addressed simultaneously, we consider the faceted particles, which yield surprisingly large scaling exponents.  While these shapes support a variety of different contact interactions, face-face contacts are presumably the most mechanically stable, and thus the most relevant in determining packing stiffness.  Yet face-face contacts produce very small changes in contact area when pressed together.  That said, it is striking that shapes with large numbers of facets have the largest $n$, i.e, behave the most nonlinearly.  A potential resolution comes from the observation that two neighboring faces, which are almost, but not exactly in parallel contact, can produce a highly non-linear dependence on the contact force if they are suddenly brought together by a small compaction.  This may also explain why increases in facet number seem to increase the scaling exponent: the probability of finding planes in near alignment should increase with the number of facets on the constituent particles
 
	In the discussion so far, by looking at the sensitivity of the effective modulus $E$ to changes in $\sigma_{con}$, we connected  sphericity $\Psi$ to the response of packings to compressive load in the limit of ${\varepsilon\to0}$.  But sphericity also affects the stress response at large $\varepsilon$, beyond yielding.  In Fig.~\ref{fig:SSAvg} the stress-strain curves for spheres, icosahedra, dodecahedra, octahedra, and also 0.92mm jacks all exhibit nearly constant stress beyond yielding, i.e.,  perfectly plastic behavior, while for the other particle geometries the stress continues to increase.  Comparison with Fig.~\ref{fig:ModScale}b shows that this change in behavior correlates not with $n$ but quite well with $\Psi$, with a cross-over value around $\Psi$ =0.85 that separates cubes from short-armed jacks.  This is consistent with the notion that shapes with extremities have a mechanism to arrest or impede large-scale plastic deformations through interlocking. However, it is more subtle in the sense that extremities have to be sufficiently pronounced to play a role and that certain convex, highly angular shapes also can mitigate large-scale plastic failure. Within our set of shapes we did not systematically explore particle aspect ratio, a factor that is likely to play an additional role, especially for cylindrical shapes. Indeed, the one fairly elongated shape, the bipyramid, despite a low sphericity ($\Psi \approx$  0.7) produces packings that yield in perfectly plastic fashion (Fig.~\ref{fig:SSAvg}a).

	\section*{Summary and Conclusions}

	This study focused on the mechanical response of granular materials of non-spherical shape. The 14 different particle geometries investigated, including convex as well as non-convex types all 3D-printed from the same material, allow us to draw a number of general conclusions that should be valid for loosely packed, random aggregates of frictional particles.  Both the effective modulus and the yield stress are found to increase with confining pressure, over the range of particles and pressures tested enabling a control of aggregate stiffness as well as strength across more than two orders of magnitude.  As a general trend, averaged over all shapes, we find that stiffness and strength are correlated. For given confinement pressure, we find that particle shape can change the effective modulus and the yield stress of the aggregate by about one order of magnitude.  This range of tunability and control is seen to depend on the confinement pressure as far as the modulus is concerned, with stronger confinement reducing the shape dependence of $E$, but is comparatively constant for $\sigma_{y}$, at least over the pressure range tested.  
	
	For each of the shapes, we find that the dependence of the aggregate stiffness on confining pressure is well described by a power law of the form  $E \propto (\sigma_{con})^n$, where the exponent $n$ encapsulates the shape dependence.  When this shape dependence is parameterized by the particle sphericity, two branches emerge. One includes the faceted, polyhedral  geometries that produce packing configurations where particles interact via several different types of contacts, in particular including those with large, rapidly varying contact area under compression. The other contains particles with arms (hexapods, dolos) and, as  limiting case of vanishing arm length, the spheres, which can each interact only via point-like contacts .  
	
	For applications our results demonstrate that granular materials hold a unique niche. Specifically, granular materials provide an extremely simple and robust solution when a material as a whole needs to transition between soft, malleable and rigid, solid-like states. This has recently become the basis for granular jamming based soft robotics applications, where highly variable compliance is achieved  by  simply changing the confinement pressure.  So far, in these applications the constituent particles have not been optimized.  Our results provide a first set of base lines in terms of the performance that can be expected from different shapes.  In particular, highly faceted polyhedral shapes appear to provide the largest range in stiffness while long-armed, interlocking shapes are least sensitive to changes in confinement.
	

    \section*{Acknowledgements}
    The authors would like to thank S. Waitukaitis and M. van Hecke for helpful discussions, as well as S. Torquato for suggesting to include experiments on truncated tetrahedra.  This work was supported by the US Army Research Office through grants no. W911NF-08-1-0140 and W911NF-12-1-0182.   For the research performed at Chicago, access to the shared experimental facilities provided by the NSF-supported Chicago MRSEC (grant no. DMR-0820054) is gratefully acknowledged.

    {\footnotesize
    \bibliography{ParticleShapeEffects}
    \bibliographystyle{rsc} 
    }


	\newpage
	\section*{Appendix}

		\subsection*{Angle of Maximum Stability}
		We define the angle of maximum stability ($\theta_m$) for our packings as the angle at which particles at the surface of a tilted bed begin to flow. To make this measurement, we poured particles into a rectangular box where they comprised a packing 10-15 particles deep. We carefully leveled the top of the packing and then slowly raised one edge of the box, while pivoting on another edge, until multiple particles began to flow down the pile. The tilt angle of the box at that point was taken as $\theta_m$.   The values listed in Table~\ref{tab:ang} are averages over 3-5 measurements for each shape. The uncertainties of $\pm$3 degrees reflect that fact that for this type of measurement the number of printed particles per shape (5000-5500) is still relatively small.

		\subsection*{Volumetric Measurements and Calculations}

		In order to determine the volumetric strain of the sample during conditioning, we started from the isotropically confined state (0.080~MPa pressure radially and axially, $q$ = 0), and surrounded the sample cell with a  sealed chamber that is filled with water. Monitoring the volume of water in that chamber as the sample is compressed provides a direct measure of changes in the sample volume, $V$.   From the measured weight of displaced water, $\Delta W(q)$,  as a function of deviatoric stress, $q$, imposed by the Instron,  the corresponding volume, $\Delta V(q) = \Delta W(q)/\rho g$, is obtained using the density of water $\rho$ and the acceleration due to gravity $g$. Because the loading piston enters the chamber as the sample is compressed, its volume $V_{piston}(q)$ needs to be  subtracted. The relative change in sample volume is then $$\frac{\Delta V(q)}{V} = \frac{1}{V}\left(\frac{\Delta W(q)}{\rho g} - V_{piston}(q)\right).$$
	
		At the end of the conditioning by cyclic loading, the reference state ($\varepsilon = 0$) is obtained by returning the load to zero. Thus, the above equation needs to be used with values for the displaced water weight and piston depth corresponding to $q$ = 0 after $N$ cycles.  From the average $\phi_0$ and the calculated volumetric changes, the average packing fraction after cycling, $\phi_{cyc}$  as listed in Table 1 is then found via $$\phi_{cyc} = \frac{\phi_0}{1 + \frac{\Delta V}{V}}.$$

		\newpage

		\subsection*{Particle Geometry}
		
		Table~\ref{tab:geom} provides the particle volume and surface area for all shapes tested in this study. The geometric quantities were calculated from the .STL models.

		\begin{table}[!h]
	      \centering
	  	\renewcommand{\arraystretch}{1.5}
	  	\renewcommand{\tabcolsep}{0.2cm} 
	  	\begin{tabular}{|l|c|c|}
	  		\hline
	  		Shape & $V_p$ (mm${}^3$)  & $A_p$ (mm${}^2$)\\
	  		\hline
	  		Spheres 		& 22.73 & 38.85 \\
	  		Tetrahedra      & 22.50 & 57.43 \\
			Cubes           & 22.50 & 47.82 \\
			Octahedra	    & 22.50 & 45.58 \\
			Dodecahedra	    & 22.50 & 42.33 \\
			Icosahedra 	    & 22.50 & 41.03 \\
			Trunc. Tetra.	& 22.50 & 49.71 \\
			Tri. Bipyr.		& 22.50 & 54.27 \\
			Tet. Frames	    & 11.91 & 65.45 \\
	 		Dolos 		    & 188.58 & 249.53 \\
			0.9mm Jacks		& 35.59 & 59.93 \\
			1.3mm Jacks		& 44.73 & 75.21 \\
			2.6mm Jacks    	& 75.98 & 129.25 \\
			3.6mm Jacks    	& 100.03 & 171.35 \\
	  		\hline
	  	\end{tabular}
	  	\caption{Particle volume and surface area for each shape tested. }
	      \label{tab:geom}
	  	\end{table}

\end{document}